\documentclass[useAMS,usenatbib]{mn2e}

\usepackage{amsmath}
\usepackage{aas_macros}
\usepackage{graphicx}

\newcommand{\CN}[1]{\texttt{#1}}
\newcommand{\model}[1]{\texttt{#1}}
\newcommand{\Jahr}{\mathrm{yr}}
\newcommand{\subpanel}[1]{\textit{#1}}

\newcommand{\VelaJr}{RX J0852.0-4622}

\newcommand{\secref}[1]{Sect.\,\ref{#1}}
\newcommand{\tabref}[1]{Tab.\,\ref{#1}}
\newcommand{\figref}[1]{Fig.\,\ref{#1}}

\newcommand{\zehn}[1]{10^{#1}}
\newcommand{\zehnh}[2]{{#1} \times 10^{#2}}

\newcommand{\msun}{M_{\odot}}

\newcommand{\km}{\textrm{km}}
\newcommand{\cm}{\textrm{cm}}

\newcommand{\erg}{\textrm{erg}}

\newcommand{\sek}{\textrm{s}}
\newcommand{\ccm}{\textrm{cm}^{3}}
\newcommand{\iccm}{\textrm{cm}^{-3}}

\begin{document}

\title[Simulations of the Vela Jr.~SNR]{Hydrodynamic simulations of the
  interaction of supernova shock waves with a clumpy environment: the
  case of the \VelaJr~(Vela Jr.) supernova remnant}

\author[M.~Obergaulinger et al.]{%%
  M.~Obergaulinger$^1$, A.F.~Iyudin$^{2,3}$, E.~M{\"u}ller$^{4}$, G.F.~Smoot$^{2,5}$
  \\
  $^1$ Departament d{\'{}}Astronomia i Astrof{\'i}sica, Universitat de
  Val{\`e}ncia, \\ Edifici d{\'{}}Investigaci{\'o} Jeroni Munyoz, C/
  Dr.~Moliner, 50, E-46100 Burjassot (Val{\`e}ncia), Spain 
  \\ $^2$ Extreme Universe Laboratory, Skobeltsyn Institute of Nuclear 
  Physics, Moscow State University by M.V. Lomonosov, \\ Leninskie Gory,
  119991 Moscow, Russian Federation
  \\ $^3$ Max-Planck-Institut f{\"u}r Extraterrestrische Physik, Postfach 1312 D-85741 Garching, Germany
  \\ $^4$ Max-Planck-Institut f{\"u}r Astrophysik, Karl-Schwarzschild-Str. 1, D-85748 Garching, Germany
  \\ $^5$ Lawrence Berkeley National Lab, 1 Cyclotron Road, Berkeley, CA 94720, USA
}

\maketitle

\begin{abstract}
  Observations in all electromagnetic bands show that many supernova
  remnants (SNRs) have a very aspheric shape.  This can be the result
  of asymmetries in the supernova explosion and of a clumpy
  circum-stellar medium.  We study the generation of inhomogeneities
  and the mixing of elements due to these two sources in
  multi-dimensional hydrodynamic simulations of the propagation of a
  supernova blast wave into a cloudy environment.  We model a specific
  SNR, VelaJr (\VelaJr).  By comparing our results with recent
  observations, we can constrain the properties of the explosion.  We
  find that a very energetic explosion of several
  $10^{51}\,\mathrm{erg}$ occurring roughly about 800 years ago as
  well as a supernova with an energy closer to the canonic value of
  $10^{51} \, \mathrm{erg}$ a few 1000 years ago are consistent with
  the shape and emission of the SNR.
\end{abstract}

\begin{keywords}
  ISM: supernova remnants%%, (stars:) supernovae: individual:
\end{keywords}

\section{Introduction}
\label{Sek:Intro}

The emission of photons across all wavelengths from many supernova
remnants (SNRs) reveals a complex geometry of these
objects. Observations indicate global asymmetries
\citep{Manchester__1987__aap__Theradiostructureofsupernovaremnants,Kesteven_Caswell__1987__aap__Barrel-shapedsupernovaremnants,Gaensler__1998__apj__TheNatureofBilateralSupernovaRemnants}
as well as a small-scale clumpy structure of many remnants
\citep[e.g.,][]{Park_et_al__2004__apj__AChandraViewoftheMorphologicalandSpectralEvolutionofSupernovaRemnant1987A,Fesen_et_al__2006__apj__TheExpansionAsymmetryandAgeoftheCassiopeiaASupernovaRemnant,Iyudin_et_al__2007__ESASP__MultiwavelengthAppearanceofVelaJr.:IsituptoExpectations}. The
observed asymmetries of SNRs can result from asymmetries imprinted in
the matter already during the supernova explosion, or they can be
created when the shock wave propagates into an inhomogeneous
interstellar medium (ISM), see, for example,
\cite{Bisnovatyi-Kogan_Silich__1995__ReviewsofModernPhysics__Shock-wavepropagationinthenonuniforminterstellarmedium}
for theoretical considerations on asymmetric appearances of SNRs, and
\cite{Gaensler__1998__apj__TheNatureofBilateralSupernovaRemnants,Wang_et_al__2002__TheMessenger__SupernovapolarimetrywiththeVLT:lessonsfromasymmetry}
on the observational interpretation of some asymmetrical SNRs.

The exact nature of the explosion mechanism is not fully understood,
but the presence of non-spherical hydrodynamic flows is a common
feature of currently discussed explosion mechanisms.  Such flows may
result, e.g., from convection and the standing accretion shock
instability (SASI), both of which play an important role for
explosions powered by the heating of matter in the stellar core by
neutrinos \citep[see, e.g.,][]{Janka_etal__2007__PRD__SN_theory}, or
from global asymmetries caused by rapid rotation of the core, in
particular in the presence of strong magnetic fields
\citep{Bisnovatyi-Kogan_Moiseenko__2008__ProgressofTheoreticalPhysicsSupplement__Core-CollapseSupernovae:MagnetorotationalExplosionsandJetFormation}.
During its non-steady propagation through the onion-shell like
structure of the stellar mantle and envelope, the supernova blast wave
is subject to Rayleigh-Taylor instabilities leading to clumpy ejecta
and an efficient large-scale mixing of the star's chemical composition
\citep{Kifonidis_et_al__2003__aap__Non-sphericalcorecollapsesupernovae.I.Neutrino-drivenconvectionRayleigh-Taylorinstabilitiesandtheformationandpropagationofmetalclumps,Kifonidis_et_al__2006__aap__Non-spherical_core_collapse_supernovae.II,Hammer_et_al__2010__apj__Three-dimensionalSimulationsofMixingInstabilitiesinSupernovaExplosions,Wongwathanarat_et_al__2010__apjl__Hydrodynamical_Neutron_Star_Kicks_in_3d}.

The interstellar medium surrounding the explosion site represents a
second source of inhomogeneities in SNRs.  A non-uniform density of
the environment can distort the shock wave even when the explosion
itself is perfectly spherically symmetric
\citep{Bisnovatyi-Kogan_et_al__1990__apss__Barrel-likesupernovaremnants}.
Various factors are responsible for shaping the spatial structure of
the ISM.  Most massive stars loose mass through an intense stellar
wind that may possess a clumpy structure, or e.g., for rapidly
rotating stars, a large-scale anisotropy.  Furthermore, massive stars
typically are born and evolve in regions of active star formation,
where the winds of many main-sequence stars and the ejecta of many
previous SNe collide forming dense clouds in the ISM.

In the following we will describe a series of three-dimensional
hydrodynamic simulations of the propagation of SN shock waves into an
inhomogeneous environment.  Apart from studying the effects that
modify the global appearance of SNRs in general, our models focus on
one particular remnant, Vela\,Jr.\ (SNR \VelaJr).  Observations of the
environment of this remnant, which have been made in different
electromagnetic bands
\citep{Dubner_et_al__1992__aaps__AstudyoftheneutralhydrogenindirectiontotheGUMnebula,
  Zinchenko_et_al__1995__aaps__Studiesofdensemolecularcoresinregionsofmassivestarformation.II.CSJ,
  Dubner_et_al__1998__aj__NeutralHydrogenintheDirectionoftheVELASupernovaRemnant,
  Yamaguchi_et_al__1999__pasj__DistributionandKinematicsoftheMolecularCloudsintheGumNebula,
  Moriguchi_et_al__2001__pasj__ACOSurveyofMolecularCloudstowardtheVelaSupernovaRemnantwithNANTEN,
  Testori__2006__aap__AradiocontinuumandneutralhydrogencounterparttotheIRASVelashell,
  Pettersson__2008__YoungStarsandDustCloudsinPuppisandVela}, reveal
the presence of a number of large clouds of atomic hydrogen (H\,I)
close to the current location of the blast wave. They also show an
enhanced emission, where the blast wave hits these clouds.  While the
observations provide valuable insights in the properties of the ISM,
they do not allow for an unambiguous determination of the properties
of the explosion.  Because the mass and energy of the ejecta are not
well determined by the observations, it is unknown whether we are
dealing with the remnant of an ordinary supernova or a more energetic
and possibly also more asymmetric hypernova event
\citep{Maeda_Nomoto__2003__apj__Bipolar_SN:Nucleosynthesis_and_Implications_for_Abundances_in_EMP_Stars}.

Age estimates for Vela\,Jr.\ range from $\sim 700\,$yrs
\citep{Iyudin_et_al__2005__aap__XMM-NewtonobservationsofthesupernovaremnantRXJ0852.0-4622_GROJ0852-4642}
up to a few $1000\,$yrs
\citep{Katsuda_et_al__2009__AdvancesinSpaceResearch__IsVelaJrayoungsupernovaremnant}.
The question of the age of the SNR is closely linked to the
determination of its distance, for which a similar range of values has
been obtained.  The distance may be as large as 750 pc, as suggested
by
\cite{Katsuda__2008__apjl__TheSlowX-RayExpansionoftheNorthwesternRimoftheSupernovaRemnantRXJ0852.0-4622}
based on an estimate of the expansion rate of the SNR.
\cite{Aschenbach_et_al__1999__aap__ConstraintsofagedistanceandprogenitorofthesupernovaremnantRXJ0852.0-4622_GROJ0852-4642},
analysing X-ray and $\gamma$-ray data, argued for a much lower
distance of around 200 pc.
\cite{Iyudin_et_al__2005__aap__XMM-NewtonobservationsofthesupernovaremnantRXJ0852.0-4622_GROJ0852-4642}
present arguments for a distance at the lower end of the range of
values, among these a possible spatial correlation between and the
Vela Jr.~SNR and other objects in the Vela region
\citep{Redman_et_al__2002__mnras__KinematicsofthePencilnebula(RCW37)anditsassociationwiththeyoungVelasupernovaremnantRXJ0852.0-4622,Pozzo_et_al__2000__mnras__Thediscoveryofalow-masspre-main-sequencestellarassociationaround$gamma$Velorum,Woermann_et_al__2001__mnras__KinematicsoftheGumnebularegion}.
\cite{Kim_et_al__2012__apj__Far-ultravioletSpectralImagesoftheVelaSupernovaRemnant:SupplementsandComparisonswithOtherWavelengthImages}
based on UV measurements of Vela SNR that include also Vela Jr.,
region E of
\cite{Kim_et_al__2012__apj__Far-ultravioletSpectralImagesoftheVelaSupernovaRemnant:SupplementsandComparisonswithOtherWavelengthImages}
give arguments for a smaller distance to this remnant.  If these
arguments hold, it is possible to use the distance of the Vela SNR as
a good estimate for the distance of Vela Jr., placing it at a distance
between 300 and 400 pc.  We treat the distance and age as unknown
parameters and perform simulations for high and low distances.

To test the compatibility of various parametrized explosion models
with the observational evidence, we have set up a model for the ISM in
the Vela\,Jr.\ region placing SN ejecta of a given mass and energy at
its centre.  We then simulate with a multi-dimensional hydrodynamic
code the propagation of the shock wave into the ISM and the expansion
of the ejecta.

Our ISM model consists of a spatial distribution of overdense
(relative to the ambient medium) clouds similar to that of the clouds
inferred from observations of the Vela\,Jr.\ region.  Concerning the
latter, we note that the X-ray image of Vela\,Jr.\ might be quite
misleading. Although appearing as a proper ellipse in projection
\citep[see Figs.~1
  in][]{Aschenbach_et_al__1999__aap__ConstraintsofagedistanceandprogenitorofthesupernovaremnantRXJ0852.0-4622_GROJ0852-4642,
  Iyudin_et_al__2007__ESASP__MultiwavelengthAppearanceofVelaJr.:IsituptoExpectations},
the
X-ray bright parts of the Vela\,Jr.\ rim, particular the NW part, are
most probably caused by a strong interaction of the forward shock with
a H\,I cloud located in that region \citep[see Fig.~3
  in]{Iyudin_et_al__2007__ESASP__MultiwavelengthAppearanceofVelaJr.:IsituptoExpectations},
The properties of the H\,I cloud can be inferred from those given in
\cite{Dubner_et_al__1998__aj__NeutralHydrogenintheDirectionoftheVELASupernovaRemnant},
and its geometry. Taking into account projection effects, this
provides an estimate of $M \sim 10^4\, \msun$ for the cloud mass, and
a range of $n=10^{2} \ldots 10^{3}\, \cm^{-3}$ for the gas
density. Together with the ejecta density of $0.06 \, \cm^{-3}$
\citep{Aschenbach_et_al__1999__aap__ConstraintsofagedistanceandprogenitorofthesupernovaremnantRXJ0852.0-4622_GROJ0852-4642}
these estimates constrain the parameters used in our simulations of
the Vela\,Jr.\ environment.

In our study we varied the explosion parameters within the range
compatible with the observations. To this end we used masses and
energies for the ejecta given by the anisotropic SN models of
\cite{Maeda_Nomoto__2003__apj__Bipolar_SN:Nucleosynthesis_and_Implications_for_Abundances_in_EMP_Stars},
to test the possibility of a hypernova-like explosion.  For
comparison, we also simulated a "standard" supernova explosion with
conventional values for the mass and energy of the ejecta.

Only a limited number of simulations of the global structure of SNRs
have been performed until present.  Earlier
simulations assuming spherical symmetry
\citep{Itoh_Fabian__1984__mnras__TheeffectofacircumstellarmediumontheevolutionofyoungremnantsofTypeIIsupernovae,
  Itoh_Masai__1989__mnras__TheeffectofacircumstellarmediumontheX-rayemissionofyoungremnantsofTypeIIsupernovae}
provided approximate relations between shock radius and SN parameters.
However, they did not consider an inhomogeneous ISM, because this
requires multi-dimensional models.  Three-dimensional simulations
similar to ours have been performed by
\cite{Shimizu_et_al__2012__pasj__EvolutionofSupernovaRemnantsExpandingoutoftheDenseCircumstellarMatterintotheRarefiedInterstellarMedium},
who particularly examined the effect of a stellar wind on the ISM and
the break-out of the shock from this wind.  Our approach differs from
their more general one by considering the specific cloud properties as
observed in the Vela\,Jr.\ remnant, and by the methods we employ to
construct light curves and emissivity maps from the hydrodynamic
simulation data. We compute the emission across the entire
electromagnetic band assuming a simple cooling function
\citep{Tucker_Koren__1971__apj__RadiationfromaHigh-TemperatureLow-DensityPlasma:theX-RaySpectrumoftheSolarCorona}
in a post-processing step and use tracer particles to determine the
mixing of chemical elements in the ejecta.

Important aspects of the evolution of a SNR include the instabilities
and turbulence developing in the flow past dense clouds, and the
heating and subsequent evaporation of clouds hit by the shock wave.
While only few \emph{global} simulations of the dynamics of SNRs
exist, these effects have been studied in great detail using
high-resolution simulations of smaller sub-regions. However, these
\emph{local} simulations do not discuss the complex gobal geometry of
the SNR.  We refer the reader, in particular, to the hydrodynamic
models of
\cite{Cowie_et_al__1981__apj__SNR_evolution_in_an_inhomogeneous_medium.I-Numerical_models,
  McKee_Cowie__1977__apj__The_evaporation_of_spherical_clouds_in_a_hot_gas.II-Effects_of_radiation,
  Klein_et_al__1994__apj__Onthehydrodynamicinteractionofshockwaveswithinterstellarclouds.1:Nonradiativeshocksinsmallclouds,
  Nakamura_et_al__2006__apjs__OntheHydrodynamicInteractionofShockWaveswithInterstellarClouds.II.TheEffectofSmoothCloudBoundariesonCloudDestructionandCloudTurbulence,
  Pittard__2010__mnras__Theturbulentdestructionofclouds-II.Machnumberdependencemass-lossratesandtailformation,
  Pittard__2011__mnras__Tailsoftheunexpected:theinteractionofanisothermalshellwithacloud}.
The possible influence of magnetic fields, which we neglect in our
study, has been considered in boundaries of local models, e.g., by
\cite{MacLow_et_al__1994__apj__Shockinteractionswithmagnetizedinterstellarclouds.1:Steadyshockshittingnonradiativeclouds,Inoue_et_al__2009__apj__TurbulenceandMagneticFieldAmplificationinSupernovaRemnants:InteractionsBetweenaStrongShockWaveandMultiphaseInterstellarMedium,Inoue_et_al__2010__apjl__Two-stepAccelerationModelofCosmicRaysatMiddle-agedSupernovaRemnants:UniversalityinSecondaryShocks,Inoue_et_al__2012__apj__TowardUnderstandingtheCosmic-RayAccelerationatYoungSupernovaRemnantsInteractingwithInterstellarClouds:PossibleApplicationstoRXJ1713.7-3946}
and in global models, albeit only with two-dimensional models, by
\cite{Fang__2012__mnras__Two-dimensionalmagnetohydrodynamicssimulationsofyoungTypeIasupernovaremnants}.

The article is organised as follows: we present an overview of our
numerical method, and a description of the physical assumptions and
the initial data in \secref{Sek:MI}. We discuss the simulation results
in \secref{Sek:Results}, and summarise our findings and draw some
conclusions in \secref{Sek:SumCon}.

\section{Methods and initial data}
\label{Sek:MI}

We used the high-resolution finite-volume code AENUS \citep{Obergaulinger__2008__PhD__RMHD}
to perform a set of three-dimensional hydrodynamic simulations.  The
code employs high-order reconstruction methods (the simulations
presented here were performed with a $5^\mathrm{th}$-order method),
\emph{monotonicity-preserving}
\citep{Suresh_Huynh__1997__JCP__MP-schemes}, and approximate HLL
Riemann solvers in the multi-stage (\emph{MUSTA}) framework
\citep{Toro_Titarev__2006__JCP__MUSTA}.  We assume that the matter is
described well by a simple ideal-gas equation of state, and take into
account neither ionisation nor heat conduction.  The simulations were
performed on a cubic Cartesian domain of $40\,\mathrm{pc}$ size with
open boundaries in all directions.  The coordinate system is oriented
in such a way that the $z$-axis points towards Earth.  The
computational domain is covered by a numerical grid of $192^3$ zones.

\begin{figure}
  \centering
  \includegraphics[width=9.0cm]{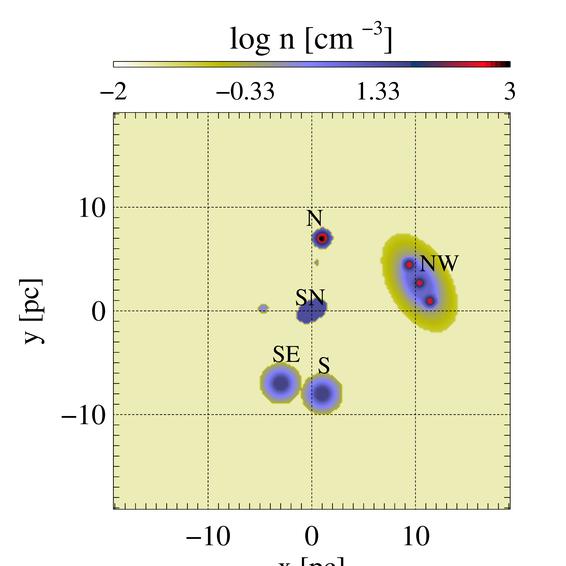}
  \caption{Colour-coded logarithm of the gas number density in a slice
    through the midplane $z = 0$ of the computational domain at the
    initial time for one of our models.  ``SN'' denotes the site of
    the explosion which is aspherical in the displayed model. ``N'',
    ``NW'', ``S'', and ``SE'' denote the large clouds modelled after
    observations of the Vela\,Jr.\ supernova remnant. }
  \label{Fig:init}
\end{figure}

\begin{table}
  \centering
  \begin{tabular}{c|rr|l|r}
    \hline
    cloud & $x$ [pc] & $y$ [pc] & axes [pc] & $n \, [1/cm^3]$ 
    \\ 
    \hline
    \CN{NW} & 10.4 & 2.7 & 5.2, 2.7 & 10
    \\
            & 9.4 & 4.45 & 0.8 & 400
    \\
            & 10.4 & 2.7 & 0.8 & 400
    \\
            & 11.4 & 0.95 & 0.8 & 400
    \\ 
    \CN{N} & 1.0 & 7.0 & 1.0 & 1000
    \\
    \CN{S} & 1.0 & -8.0 & 2.0 & 50
    \\
    \CN{SE} & -3.0 & -7.0 & 2.0 & 50
    \\
    \hline
  \end{tabular}

  \caption{Properties of the large clouds placed into the SNR
    environment which we assume to be
    of spheroidal or spherical shape.  The columns give the cloud
    name, the position of its centre (note that all four clouds are
    located in the midplane, $z=0$), the semi-major and semi-minor
    axes of the spheroid (for spherical clouds we list only one
    number, the radius), and the number density of the cloud.  The
    spheroidal cloud \CN{NW} contains three dense cores whose
    properties are listed below those of the cloud \CN{NW} itself.
  }
  \label{Tab:cloud-init}
\end{table}

We approximate the clumpy environment of the Vela\,Jr.\ supernova
explosion site by considering two populations of dense clouds,
which are embedded into an ambient ISM that has a constant gas number
density $n_{\mathrm{ISM}}$ and temperature $T_\mathrm{ISM}$.  For most
models used in our study we assume $n_{\mathrm{ISM}} = 0.025 \, \iccm$
and $T_{\mathrm{ISM}} = 10 \, \mathrm{K}$.
  
One population of clouds is based on observations of the
Vela\,Jr.\ remnant
\citep{Dubner_et_al__1992__aaps__AstudyoftheneutralhydrogenindirectiontotheGUMnebula,
  Zinchenko_et_al__1995__aaps__Studiesofdensemolecularcoresinregionsofmassivestarformation.II.CSJ,
  Dubner_et_al__1998__aj__NeutralHydrogenintheDirectionoftheVELASupernovaRemnant,
  Yamaguchi_et_al__1999__pasj__DistributionandKinematicsoftheMolecularCloudsintheGumNebula,
  Moriguchi_et_al__2001__pasj__ACOSurveyofMolecularCloudstowardtheVelaSupernovaRemnantwithNANTEN,
  Testori__2006__aap__AradiocontinuumandneutralhydrogencounterparttotheIRASVelashell,
  Pettersson__2008__YoungStarsandDustCloudsinPuppisandVela} and
consists of four large clouds situated at a distance of roughly 10\,pc
from the explosion site.  The positions and properties of these clouds
are given in \tabref{Tab:cloud-init}.  We label the four large clouds
by their location with respect to the explosion site
\figref{Fig:init}: cloud \CN{NW}, the largest one, consists of three
dense cores contained in a larger, but less dense spheroidal structure.
The two clouds \CN{S} and \CN{SE} are identical spheres, and cloud
\CN{N}, the densest one, is of spherical shape, too.  To convert the
observed positions of the clouds on the sky and their observed angular
diameters into physical lengths, we had to assume a distance to the
SNR, which we set equal to 300\,pc \citep[see discussion about the
  distance to
  Vela\,Jr.\ in][]{Iyudin_et_al__2007__ESASP__MultiwavelengthAppearanceofVelaJr.:IsituptoExpectations}.
 
The second population of clouds consists of a large number of smaller
spherical members, presumably below the resolution limit of the
observations, at random positions in the sky, whose number density
$n_{\mathrm{random}} = 20 \iccm$.  The cloud radii are randomly
distributed between 0.02\,pc and 0.5\,pc. This population of clouds was inferred 
from optical observations of Vela SNR environment, see, for example Pakhomov et al. (2011) and references therein. 

All clouds are assumed to be in pressure equilibrium with the
surrounding matter, i.e, $T_{\mathrm{cloud}} = T_{\mathrm{ISM}} =
n_{\mathrm{ISM}}/ n_{\mathrm{cloud}}$.  The large spherical clouds
(\tabref{Tab:cloud-init}) have a density falling off as $r^{-2}$,
where $r$ is the distance from the cloud centre. For the spheroidal
cloud \CN{NW} the density is constant on spheroidal surfaces falling
off as $d^{-2}$, where $d$ is given by
\begin{eqnarray}
  \label{Gl:d-Ellipsoid}
  d ^2 & = &
  \frac{r^2 \sin^2 (\theta - \theta_{\mathrm{c}}) 
    \cos^2 (\phi -   \phi_{\mathrm{c}})}{a_1^2} 
  \\
  \nonumber
  & & + 
  \frac{r^2 \sin^2 (\theta - \theta_{\mathrm{c}}) 
    \sin^2 (\phi   - \phi_{\mathrm{c}})}{a_1^2}
  + 
  \frac{r^2 \cos^2 ( \theta - \theta_{\mathrm{c}})}{a_2^2}
  .
\end{eqnarray}
Here $(r,\theta,\phi)$ are the spherical coordinates of a point inside
the cloud w.r.t.\ the centre of the cloud, $\theta_{\mathrm{c}}$ and
$\phi_{\mathrm{c}}$ give the orientation of the (oblate) spheroid,
i.e., the direction of its semi-major axis $a_1$, and $a_2$ is the
spheroid's semi-minor axis.

\begin{table*}
  \centering
  \begin{tabular}{l|cc|l}
    name & $M_{\mathrm{SN}} \, [M_{\odot}]$ 
    & $E_{\mathrm{SN}} \, [\mathrm{10^{51} erg}]$ 
    & remarks
    \\
    \hline
    \model{S25A} & 6.0 & 6.7 & reference model, progenitor of 25 $\msun$
    \\
    \model{S40} & 10.2 & 10.9 & progenitor of 40 $\msun$
    \\
    \model{S25B} & 6.0 & 1.0 & weak explosion
    \\
    \model{S25BB} & 6.0 & 0.6 & weak explosion
    \\
    \model{S25As} & 6.0 & 6.7 & spherical explosion
    \\
    \model{S25Bs} & 6.0 & 1.0 & weak spherical explosion
    % \\
    % \model{S25Aw} & 6.0 & 6.7 & pulsar wind
    \\
    \hline
    \model{S25Anc} & 6.0 & 6.7 & no clouds
    \\
    \model{S25A4c} & 6.0 & 6.7 & only the four main clouds
    \\
    \model{S25AC} & 6.0 & 6.7 & 1200 random clouds of $n_{\mathrm{cl}}
    = 20 \, \iccm$
    \\
    \model{S25Aw} & 6.0 & 6.7 & warm ISM, $T_{\mathrm{ISM}} = 1000 \, \mathrm{K}$
    \\
    \model{S25Ad0.1} & 6.0 & 6.7 & dense ISM, $\rho_{\mathrm{ISM}} = 0.1
    \, \iccm$
    \\
    \model{S25Ad1.0} & 6.0 & 6.7 & dense ISM, $\rho_{\mathrm{ISM}} = 1.0
    \, \iccm$
    \\
    \model{S25Ad10} & 6.0 & 6.7 & dense ISM, $\rho_{\mathrm{ISM}} = 10
    \, \iccm$
    \\
    \hline
    \model{S25AD} & 6.0 & 6.7 & SNR placed at a distance of $D = 750 \,\mathrm{pc}$
    \\
    \hline
  \end{tabular}
  \caption{Overview of the computed models: The first column displays
    the model name, the second the mass of the ejecta, and the third
    the explosion energy.  Further properties of the models are given
    in the last column.
  }
  \label{Tab:init-expl}
\end{table*}

We do not take into account the effects of gravity in our simulations,
since gravity does not affect the evolution of the SNR, because the
gravitational binding energy of the gas comprises only a small
fraction of its total energy.  Moreover, the gravitational time scale,
i.e. the free-fall time scale of the unshocked ISM, is longer than the
time scales we are interested in.

We initiate the explosion by placing gas of a total mass
$M_{\mathrm{SN}}$ and energy $E_\mathrm{SN}$ at the centre of the
computational grid, and assume that the ejecta have the shape of a
prolate spheroid of semi-major axis $A_{\mathrm{SN}}$ and semi-minor
axis $a_{\mathrm{SN}}$, or of a sphere of radius $R_{\mathrm{SN}}$.
All three lengths are in the range of order 1 to 2\,pc, i.e., the
initial explosion site is covered by 10 to 20 grid zones per spatial
dimension, only.  The imposed explosions are inspired by models 25A,
25B, and 40A of
\cite{Maeda_Nomoto__2003__apj__Bipolar_SN:Nucleosynthesis_and_Implications_for_Abundances_in_EMP_Stars}.
corresponding to progenitors of 25 and 40 solar masses, respectively.
The ejecta masses $M_{\mathrm{SN}}$ and the explosion energies
$E_\mathrm{SN}$ of our models are given in \tabref{Tab:init-expl}.  As
our grid resolution is too coarse to allow for a direct mapping of the
\cite{Maeda_Nomoto__2003__apj__Bipolar_SN:Nucleosynthesis_and_Implications_for_Abundances_in_EMP_Stars}
models, we had to rely on ad-hoc profiles for the hydrodynamic
variables of the initial ejecta.  For simplicity, we assume that the
total explosion energy is composed of 5\% thermal energy and 95\%
kinetic energy, the density and pressure of the ejecta are uniform,
and the initial velocity is purely radial and varies linearly with
spheroidal distance $d$ (see \eqref{Gl:d-Ellipsoid}).

In the simulations we would also like to follow the propagation of
small clumps of different elements ejected in the explosion through
the ISM.  However, because the sizes of these clumps are far below the
affordable grid resolution, we model these clumps as test particles
that are passively advected with the flow, i.e. the particle positions
are determined by solving the equation of motion,
\begin{equation}
  \label{Gl:Testp}
  \frac{\partial \vec r_{\mathrm{p}}}{\partial t} 
  = \vec v_{\mathrm{flow}} (\vec r_{\mathrm{p}}),
\end{equation}
where $\vec r_{\mathrm{p}} $ and $ \vec v_{\mathrm{flow}} (\vec
r_{\mathrm{p}})$ are the time-dependent position of a particle and the
flow velocity at that position, respectively. In the initial models,
we place a number of particles at random positions near the explosion
site and assign some elemental distribution to each of it.  By
following their advection, we are able to determine the spatial
distribution of elements in the SNR.  We distinguish two populations
of tracer particles, which we call the \emph{outflow} component and
the \emph{spherical} component.  The outflow component is composed of
all particles initially situated in a cone of opening angle
$30^{\circ}$ around the semi-major axis of the prolate explosion
spheroid. The remaining particles make up the spherical component.  In
our simulations we follow the evolution of the elemental abundances of
oxygen, titanium, and nickel.

The collision rates between ions and electrons are very low in the
tenuous gas inside the Vela Jr. SNR, typical ionisation ages being
around $10^{10} \, \mathrm{s} / \mathrm{cm}^3$
\citep{Aschenbach_et_al__1999__aap__ConstraintsofagedistanceandprogenitorofthesupernovaremnantRXJ0852.0-4622_GROJ0852-4642}.
Therefore, no equilibrium between ions and electrons can be
established within the evolutionary times of interest here, i.e., the
two kinds of particles are characterised by two different
temperatures, $T_{e}$ and $T_{i}$.  When a fluid element is passed by
the shock, the ions are heated, but thermal energy is transferred to
the electrons only on very long time scales by infrequent collisions.
The delayed heating of the electrons has important consequences for
the emission of photons by the gas, which is not determined by the
temperature of the ions but rather by the lower electron temperature.
Therefore, we consider electrons and ions separately in our
simulations to avoid overestimating the luminosity of the SNR.

In our treatment of the \emph{non-equilibrium ionisation} (NEI), we
use an expression for the time scale of equilibration between
electrons and ions of mass $m_j$ and charge $Z_j$ given by
\cite{Spitzer__1962__PhysicsofFullyIonizedGases},
\begin{equation}
  \label{Gl:Spitzer-teq}
  t_{\mathrm{eq; j}} = \frac{3 m_j m_e k^{3/2}}{ 8 \sqrt{2 \pi}
    n_j Z_j ^2 Z_e ^2 e ^ 4 \ln \Lambda } \left( \frac{T_e}{m_e} +
    \frac{T_{\mathrm{ion}}}{m_j}\right)^{3/2}.
\end{equation}
Here, $m_e$ and $Z_e$ are the mass and charge of the electron,
respectively (charges are measured in units of the elementary charge,
$e$); $n_j$ is the number density of the ions, and $\ln \Lambda$ is
the Coulomb logarithm.  Using this time scale, the equation for the
internal-energy density of the electron gas, $\varepsilon_e$, reads
\begin{equation}
  \label{Gl:Spitzer-Ee}
  \partial_{t} \varepsilon_e + \vec \nabla ( \varepsilon_e \vec v ) 
  = - P_e \vec \nabla \cdot \vec v + Q,
\end{equation}
where the electron pressure is derived from the internal energy by
$P_e = \varepsilon_e (\gamma_e - 1)$ with $\gamma_e = 5/3$, and the
equilibration source term $Q = 3/2 \, n_e k ( T_{\mathrm{ion}} - T_e )
\sum_j t_{\mathrm{eq; j}} ^ {-1}$.
\\
Since the most detailed treatment, i.e., evolving equations governing
the advection, ionisation and recombination of a large set of elements
in all relevant ionisation states, is beyond the scope of this
article, we used a simplified approach replacing the set by only one
representative ion.  We set $m_j = 1.25 \, \mathrm{amu}$ and $Z_j =
1$, i.e., appropriate for a mixture of hydrogen and helium.  As a
further simplification, we use a constant Coulomb logarithm, $\ln
\Lambda = 33.9$.  While such an approach is inferior to a detailed NEI
simulation, it allows us to capture the essential NEI effects with an
acceptable uncertainty when compared to the uncertainties due to the
unknown parameters of the explosion and the ISM.

Our hydrodynamic simulations provide the time-dependent
three-dimensional distributions of ejecta parameters, like .e.g,
density or temperature.  To compare these results with the observed
emission of SNR, one has to couple the hydrodynamic evolution of the
matter to the radiative transfer of photons emitted, absorbed, and
scattered by the gas.  Fortunately, the properties of the gas allow
for several simplifications to this otherwise very complex task.
First and foremost, we note that the gas is optically thin to the
radiation we are interested in, i.e., X-ray and UV photons.  Moreover,
radiation carries away only a small fraction of the total energy of
the expanding ejecta.  This allows us to determine the emission by
post-processing the hydrodynamic simulation results.

Given the hydrodynamic nature of our models, we compute the spectral
emissivity $\frac{\mathrm{d} \epsilon_{B}}{\mathrm{d} \lambda}$
of thermal bremsstrahlung according to a simplified version of
the expression given by
\cite{Tucker_Koren__1971__apj__RadiationfromaHigh-TemperatureLow-DensityPlasma:theX-RaySpectrumoftheSolarCorona},
where
\begin{equation}
  \label{Gl:Brems-emis}
  \frac{\mathrm{d} \epsilon_{B}}{\mathrm{d} \lambda} =
  \frac{\zehnh{2.04}{22} \left[\frac{\mathrm{erg}}{\mathrm{cm^3 \, s
          \, \AA}}\right] }
    {\lambda^2 T_6^{1/2}} Z^2 n_e
  n_{\mathrm{ion}} g_{\mathrm{B}} 
  \exp \left(- \frac{144 \, \AA }{\lambda  T_6} \right).
\end{equation}
Here, $T_6$ is the (electron) temperature in units of $10^{6} \,
\mathrm{K}$.  Our implementation assumes only one species of ions;
furthermore, we approximate the Gaunt factor by unity.  By integrating
\eqref{Gl:Brems-emis} over wavelength between the wavelengths
corresponding to the minimum and maximum photon energies
$e_{\mathrm{min}}$ and $e_{\mathrm{max}}$, we obtain the emission in
various energy bands.  We will focus our analysis on X-ray and UV
bands.  We will present light curves of our models in a number of
intervals distributed logarithmically over photon energy
(\tabref{Tab:Energybands}).  An accurate treatment of other
  (non-thermal) processes requires a more sophisticate modelling
  approach.  Therefore, we defer such a more complete analysis of the
  photon emission to a later time.

\begin{table}
  \centering
  \begin{tabular}{lll}
    $e_{\mathrm{min}}$ [keV] & $e_{\mathrm{c}}$ [keV] & $e_{\mathrm{max}}$ [keV]
    \\
    \hline
    0.080 & 0.10 & 0.1282 
%%    \\ 0.1282 & 0.16 & 0.2055
    \\ 0.2055 & 0.26 & 0.3293
%%    \\ 0.3293 & 0.42 & 0.5278
    \\ 0.5278 & 0.67 & 0.8459
%%    \\ 0.8459 & 1.07 & 1.356
    \\ 1.356 & 1.72 & 2.173
%%    \\ 2.173 & 2.75 & 3.482
    \\ 3.482 & 4.41 & 5.581
%%    \\ 5.581 & 7.07 & 8.944
    \\ 8.944 & 11.32 & 14.33
%%    \\ 14.33 & 18.15 & 22.97
    \\ 22.97 & 29.08 & 36.82
    % \\ 36.82 & 46.61 & 59.01
    % \\ 59.01 & 74.71 & 94.57
    % \\ 94.57 & 119.73 & 151.6
    % \\ 151.6 & 191.88 & 242.9
    % \\ 242.9 & 307.53  & 389.3
    % \\ 389.3 & 492.87 & 624.0
    % \\ 624.0 & 789.91 & 1000
    \\ 0.080 & 0.10 & 0.120
    \\ 1.000 & 4.0 & 7.000
  \end{tabular}
  \caption{%%
    The minimum, mean, and maximum photon energies of the bands we
    consider for generating light curves of our models.
  }
  \label{Tab:Energybands}
\end{table}

\section{Simulations and results}
\label{Sek:Results}

In the following we will describe the most important results from our
simulations.  We will discuss the evolution of a reference model in
more detail before turning to a series of models in which we vary
subsequently different initial parameters.

\subsection{Observational facts}
\label{sSek:Obs}

\begin{figure*}
  \centering
  \includegraphics[width=6.5cm]{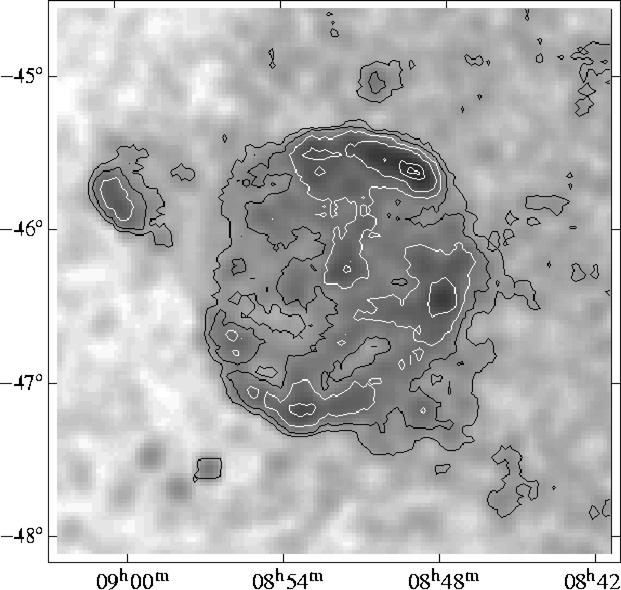}
  \includegraphics[width=10cm]{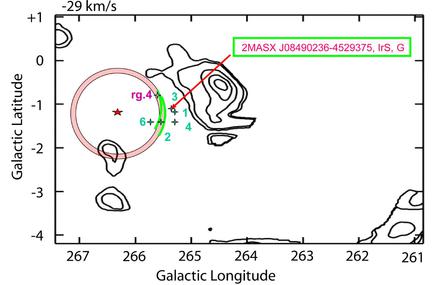}
  \caption{%%
    The \subpanel{left panel}, taken from
    \citet{Aschenbach_et_al__1999__aap__ConstraintsofagedistanceandprogenitorofthesupernovaremnantRXJ0852.0-4622_GROJ0852-4642},
    shows a ROSAT image of SNR \VelaJr~in X-rays from 1.3 keV to 2.4
    keV; coordinates are right ascension and declination of the epoch
    J2000.0.  The \subpanel{right panel} shows a schematic of the
    interaction of a cloud with the shock wave of the SNR
    \citep{Iyudin_et_al__2007__ESASP__MultiwavelengthAppearanceofVelaJr.:IsituptoExpectations}.
  }
  \label{Fig:Observations}
\end{figure*}

We briefly summarise the main observational findings which we later
compare with our numerical results:
\begin{itemize}
\item The geometry of the remnant, as observed in, e.g., X-rays and UV
  \citep[see \figref{Fig:Observations} for an example of X-ray
  observations
  by][]{Aschenbach_et_al__1999__aap__ConstraintsofagedistanceandprogenitorofthesupernovaremnantRXJ0852.0-4622_GROJ0852-4642}
  appears to be almost circular.  At the level of the brightness
  chosen in these observations, the axis ratio is close to unity, but
  varies from waveband to waveband, and with the sensitivity
  thresholds of the imaging instruments.  In addition, several large
  spots show enhanced emission, particularly near the outer border of
  the circular X-ray-emitting region.
\item An estimate of the expansion rate of the remnant of $\sim 0.023
  \% \, \Jahr^{-1}$ was obtained by
  \cite{Katsuda__2008__apjl__TheSlowX-RayExpansionoftheNorthwesternRimoftheSupernovaRemnantRXJ0852.0-4622}
  for NW X-ray bright rim of the remnant. Because the rim might be
  bright as a result of the strong interaction of the SNR blast wave
  with the HI cloud in that region \citep[][
  Fig.~3]{Iyudin_et_al__2007__ESASP__MultiwavelengthAppearanceofVelaJr.:IsituptoExpectations},
  this measurement is related to the cloud shock velocity of this
  bright spot.
  \cite{Katsuda__2008__apjl__TheSlowX-RayExpansionoftheNorthwesternRimoftheSupernovaRemnantRXJ0852.0-4622}
  agree with such an interpretation.
\item Estimates of the energy emitted in X-rays at four bright spots
  (at the centre and three along the rim of the SNR) have been
  obtained by
  \cite{Slane_et_al__2001__apj__RXJ0852.0-4622:AnotherNonthermalShell-TypeSupernovaRemnant(G266.2-1.2)}.
  To convert the fluxes to luminosities, we have to assume a distance
  to the SNR. Taking the total, absorption corrected flux of the
  high-temperature X-ray emission component of Vela Jr. as
  $F_{\mathrm{X}} (0.1-2.4 \, \mathrm{keV}) = \zehnh{3}{-10} \, \erg
  \, \mathrm{cm}^{-2} \, \mathrm{s}^{-1}$
  \citep{Aschenbach__1998__nat__Discoveryofayoungnearbysupernovaremnant}
  for $d \sim 300 \, \mathrm{pc}$, we derive the luminosity of remnant
  in the 0.1-2.4 keV energy range as $L_{\mathrm{X}}= \zehnh{3}{33} \,
  \erg \, \mathrm{s}^{-1}$.  The bright spots near the outskirts of
  the SNR have a luminosity of a few $10^{32} \, \erg$ / $\sec$ each
  \citep{Aschenbach_et_al__1999__aap__ConstraintsofagedistanceandprogenitorofthesupernovaremnantRXJ0852.0-4622_GROJ0852-4642}.
  The emission in extreme the UV is of a similar or larger magnitude
  \citep{Kim_et_al__2012__apj__Far-ultravioletSpectralImagesoftheVelaSupernovaRemnant:SupplementsandComparisonswithOtherWavelengthImages,Nishikida_et_al__2006__apjl__Far-UltravioletSpectralImagesoftheVelaSupernovaRemnant}.
  The observed spectra indicate that the emission of the SNR is
  dominated by non-thermal processes, most likely synchrotron
  radiation. Thus, estimates of the luminosity provide an upper bound
  of the thermal contribution only.  Further evidence for non-thermal
  processes is provided by observations in the TeV range
  \citep{Aharonian_et_al__2007__apj__H.E.S.S.ObservationsoftheSupernovaRemnantRXJ0852.0-4622_Shell-TypeMorphologyandSpectrumofaWidelyExtendedVeryHighEnergyGamma-RaySource}.
\item Observations
  \citep{Dubner_et_al__1998__aj__NeutralHydrogenintheDirectionoftheVELASupernovaRemnant}
  suggest the presence of large clouds of neutral hydrogen in the projected vicinity to the SNR
  \VelaJr.  The \subpanel{right panel} of
  \figref{Fig:Observations} shows a schematic of a particularly large
  cloud (NW), possibly in  interaction with the SNR shock wave
  \citep{Iyudin_et_al__2007__ESASP__MultiwavelengthAppearanceofVelaJr.:IsituptoExpectations}.
\end{itemize}

These findings show a rather complex structure of the remnant.
However, a rather high degree of uncertainty remains, and therefore,
they do not unambiguously point to a particular mechanism for creating
the inhomogeneous appearance of the SNR.  We test one possible,
viz.~the interaction of the SN blast wave with an inhomogeneous medium
containing large clouds with a large overdensity w.r.t.~the mean
density of the ISM.

A complication arises in the comparison of our simulation results with
observational data due to a potential contamination by emission from
the nearby (in projection on the celestial sphere) SNR Vela.
Furthermore, as noted above, the observed emission is non-thermal to a
high degree, i.e., only upper limits can be given for the thermal
contribution, and thermal line emission is particularly strong in the
soft X-ray band.  Our simplified analysis is based on thermal
bremsstrahlung only.  Thus, a comparison of simulation results and
observations is somewhat restricted.  Based on his Eqn.~(52),
\cite{Vink__2012__aapr__Supernovaremnants:theX-rayperspective}
estimates that non-thermal synchrotron emission requires relatively
high shock velocities of $> 2000 \km / \sec$.  Applying this criterion
to the hydrodynamic evolution of our models we can assess the
importance of non-thermal emission.  A more thorough treatment is
hindered by lack of detailed information about the magnetisation of
the ISM around the Vela\,Jr.\ supernova remnant, which can be
significantly higher behind the shock than in the ambient medium due
to efficient field amplification at the shock surface
\citep{Zirakashvili_Ptuskin__2012__AstroparticlePhysics__NumericalsimulationsofdiffusiveshockaccelerationinSNRs}.
Including a magnetic field into our simulations, while in principle
possible, would increase the parameter space considerably, in
particular if we take into account that estimates of the field
strengths vary widely.
\cite{Kishishita_et_al__2013__aap__NonthermalemissionpropertiesofthenorthwesternrimofsupernovaremnantRXJ0852.0-4622}
estimate $B \sim 6-10 \, \mu G$ for the field in the NW regions of the
SNR and $B_0 \sim 1 \, \mu G$ for the ISM field, while
\cite{Berezhko_Voelk__2010__aap__NonthermalandthermalemissionfromthesupernovaremnantRXJ1713.7-3946}
find a better agreement with observations if they assume a downstream
field of $B \sim 140 \, \mu G$, corresponding to an ambient field of
$B_0 = 25 \, \mu G$.  This uncertainty limits our ability to, e.g.,
setup the initial condition for self-consistent MHD simulations.

Because of observational uncertainties, but even more because of the
unknown initial conditions and the simplifications in our models, we
do not expect that one particular model perfectly matches the
observations.  However, we try to identify simulation results which
resemble the observations closest.

\subsection{General features}
\label{sSek:Gen}

\begin{figure*}
  \centering
  \includegraphics[width=7.4cm]{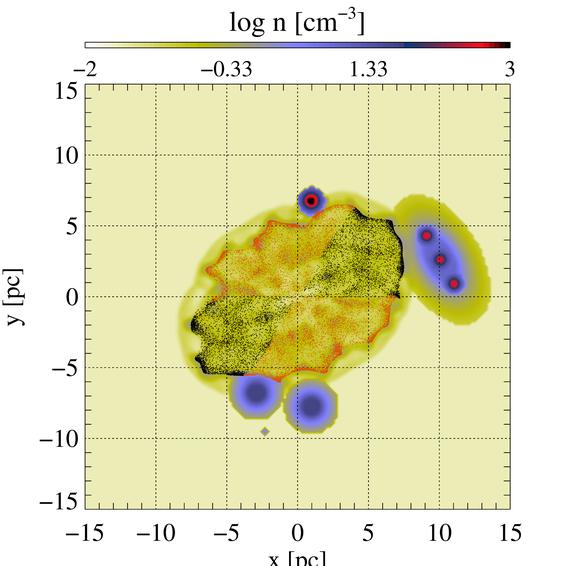}
  \includegraphics[width=7.4cm]{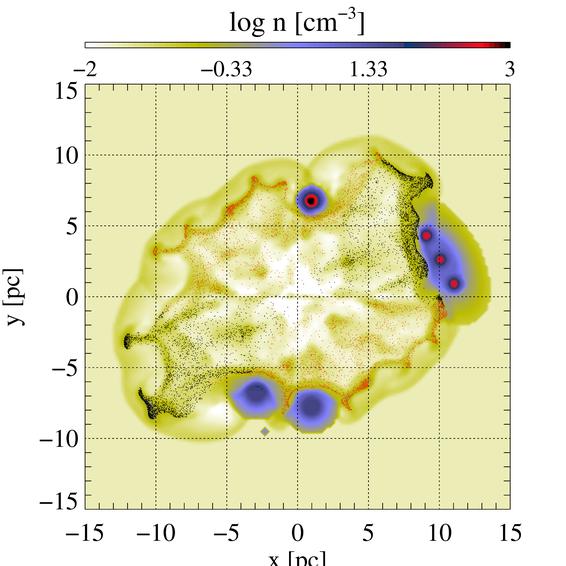}
  \includegraphics[width=7.4cm]{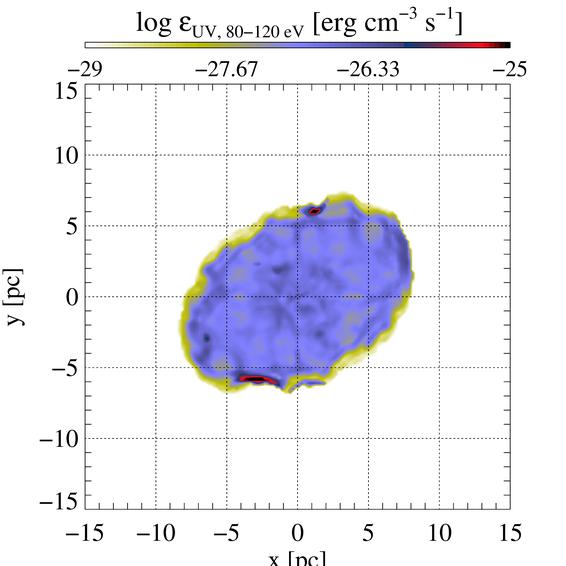}
  \includegraphics[width=7.4cm]{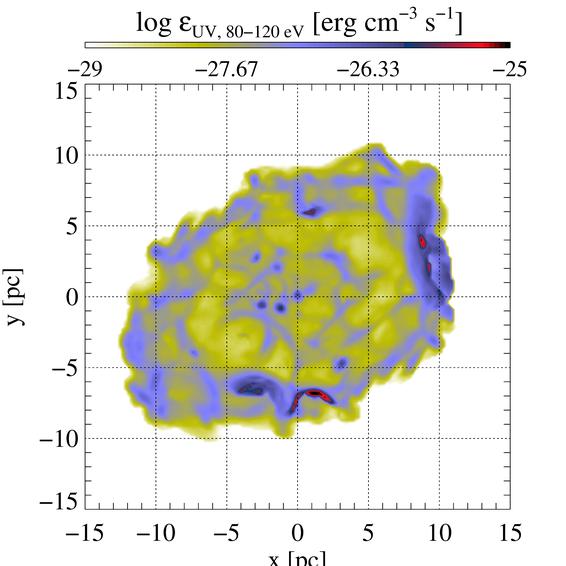}
  \includegraphics[width=7.4cm]{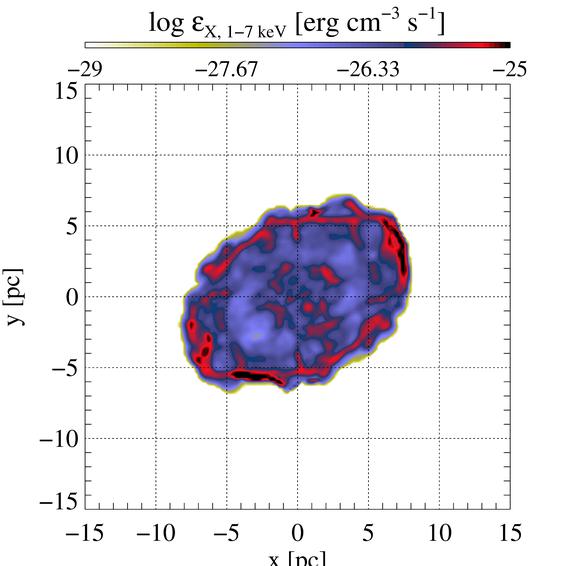}
  \includegraphics[width=7.4cm]{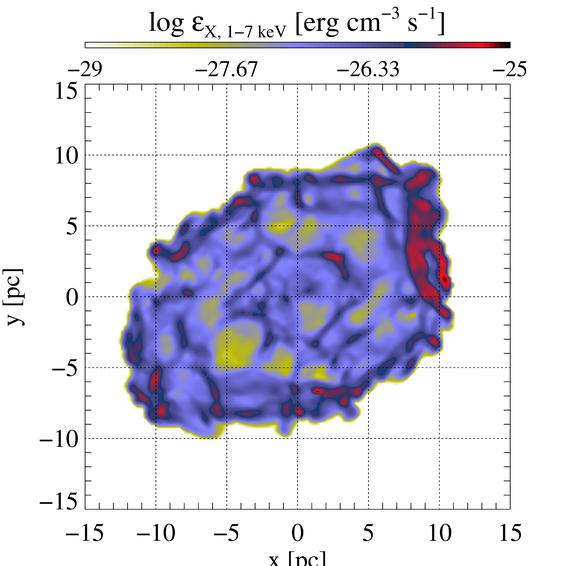}
  \caption{\subpanel{Top panels}: a slice through model \model{S25A}
    at times $t = 480\, \Jahr$ (\subpanel{left}) and $t = 900\, \Jahr$
    (\subpanel{right}) showing besides the density distribution the
    position of tracer particles close to the midplane.  Particles
    from the outflow and spherical components are shown by black and
    orange dots, respectively.
    \subpanel{Middle panels}: emissivity maps in the UV band 80 - 120 eV.
    \subpanel{Bottom panels}: emissivity maps in the X-ray band 1 - 7
    keV.  Both emissivity maps show the logarithm of 
    the cooling function integrated along the line of sight (the
    $z$-direction) through the entire computational domain.  
  }
  \label{Fig:S25A-2d-900a}
\end{figure*}

We chose a simulation with an explosion based on the SN model 25A by
\cite{Maeda_Nomoto__2003__apj__Bipolar_SN:Nucleosynthesis_and_Implications_for_Abundances_in_EMP_Stars}
as the first model we discuss.  The progenitor of this explosion is a
star of 25 solar masses exploding in a prolate geometry, releasing a
mass of $M_{\mathrm{SN}} = 6 \msun$ and an energy of $e_{\mathrm{SN}}
= 6.7 \times 10^{51}\, \mathrm{erg}$.  We approximate the aspherical
geometry of the explosion very roughly by a prolate spheroid with an
axis ratio of $1.6 \, \mathrm{pc} : 1.0 \, \mathrm{pc} : 1.0 \,
\mathrm{pc}$.  We follow the evolution of the simulated remnant till
an age of $t = 1500\, \Jahr$.

During the first $60\, \Jahr$, the explosion roughly retains its
spheroidal shape although the axis ratio decreases somewhat, i.e., the
ejecta become slightly more spherical.  This evolution, a posteriori,
justifies that we assume in our simulations only slightly spheroidal
ejecta, because the possibly more pronounced initial asphericity at
the onset of the explosion has already decreased during the time it
took the shock wave to expand to a size of $\sim 1 \, \mathrm{pc}$,
when we start our simulations.

The post-shock state consists of tenuous hot gas having a density
around $n \sim 1 \, \iccm$ and an electron temperature around $T_e
\sim 1 \, \mathrm{keV}$.  The gas is a strong emitter of mainly X-ray
photons.  The total luminosities in the X-ray and UV bands at $t = 60
\, \Jahr$ are $L_{\mathrm{X}} \sim \zehnh{8.2}{34} \, \erg/s$ and
$L_{\mathrm{EUV}} \sim \zehnh{4.9}{33} \, \erg/s$, respectively.
While these values only slightly exceed the luminosities estimated
from observations, the emission decreases quickly to a level lower
than that derived from observations. We note here that most of the EUV
luminosity of the remnant is produced by the nonradiative shocks
inside the remnant
\citep{Kim_et_al__2012__apj__Far-ultravioletSpectralImagesoftheVelaSupernovaRemnant:SupplementsandComparisonswithOtherWavelengthImages,Nishikida_et_al__2006__apjl__Far-UltravioletSpectralImagesoftheVelaSupernovaRemnant,Nichols_Slavin__2004__apj__ShockedCloudsintheVelaSupernovaRemnant}.

The spheroidal shape of the ejecta is clearly destroyed when the shock
wave hits the closest clouds (see density maps in
\figref{Fig:S25A-2d-900a}).  Though these are small, they are able to
divert the ejecta and shield the regions immediately behind them to a
certain degree.  On impact, the shock wave heats the surface of the
clouds, and the heated rather dense gas starts to emit X-ray
radiation.  This can be seen in the emissivity maps in
\figref{Fig:S25A-2d-900a}.

Upon passage of the shock wave, the ionic component of the gas is
heated to very high temperatures, but the gas starts to radiate only
after the electrons have been heated up to roughly the ion temperature
by collisions.  Since the frequency of collisions increases with
density, equilibration occurs fastest in the densest parts of the SNR.
The central regions of the SNR possess a very low density, and though
the density is considerably higher right behind the shock wave, it is
still low compared to that of the clouds.  Consequently, once hit by
the shock, the combination of high (electron) temperatures and
densities makes the clouds prominent sources of radiation.  Thus, the
radiation emitted by the SNR depends very strongly on the properties
of the clouds.

The radiative energy loss is small, i.e., the gas remains hot and
therefore radiates for a long time. Clouds can remain strong emitters
of radiation throughout the entire evolution unless they are dispersed
by the impact.  We find a notable disruption only for clouds of very
small radius or low density.  Clouds above a radius of $\sim 2 \,
\mathrm{pc}$ remain intact, except for some stripping of surface
material and some deformations.  The clouds which retain their
position, size and shape, act merely as obstacles for the blast wave.
This behavior holds even more for the few massive clouds \CN{NW},
\CN{N}, \CN{S}, and \CN{SE}, respectively.  Neither does the shock
wave disrupt any of the clouds nor is it able to modify the
substructure of the cloud \CN{NW}.  Because of their high densities,
the four large clouds are the strongest emitters of radiation in the
SNR.

\begin{figure}
  \centering
  \includegraphics[width=9cm]{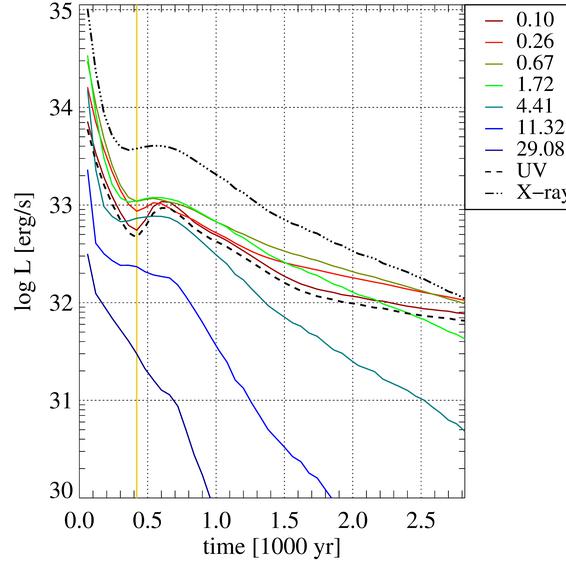}
  \caption{%%
    Light curves of model \model{S25A}.  The solid lines show the
    luminosity of the SNR model in different UV and X-ray bands (see
    \tabref{Tab:Energybands}); the legend indicated the energy of the
    band in keV.  The black dashed and dash-dot-dot-dotted lines are
    the luminosities in the bands of $80-120$ eV and $1-7$ keV,
    respectively.  The vertical line marks the beginning of the
    interaction between the shock wave and the clouds.
  }
  \label{Fig:S25A-LC}
\end{figure}

As the shock wave expands, the density and temperature of the
post-shock gas decrease, i.e., the luminosity decreases.  We show the
light curves in several bands in UV and X-ray for model \model{S25A}
in \figref{Fig:S25A-LC}.  At all times, the emission is dominated by
photon energies around $\sim 1 \, \mathrm{keV}$ (green and blue solid
lines).  For this model, the X-ray emission between 1 and 7 keV
(dash-dot-dot-dotted black line) exceeds the emission in the extreme
UV band between 80 and 120 eV (dashed black line) by a factor $\sim 5$
throughout most of the evolution.  The X-ray emission exhibits a
plateau at a total luminosity $L_{\mathrm{X}} \sim \zehn{34} \,
\erg/\sek$ before it decreases gradually, while the UV luminosity
begins to increase at around $t \sim 480 \, \Jahr$, when the blast
wave starts to impact the main clouds (the beginning of the
interaction is marked by a vertical line in the figure).  The X-ray
emission peaks at $t \sim 600 \, \Jahr$, possibly facilitating
detection of the SNR in X-rays.  At $t = 900 \, \Jahr$ the shock-cloud
interaction is in full swing, and the clouds are being hit by the
ejecta (cf.~the positions of the tracer particles in the \subpanel{top
  right panel} of \figref{Fig:S25A-2d-900a}).  At this point, the flow
geometry is already quite complex, leading to mixing in the gas.  In
both bands, several single bright spots can be identified that
coincide with the clouds.  In particular, we find a strong brightening
of the NW limb of the SNR (see the \subpanel{middle and bottom} panels
of \figref{Fig:S25A-2d-900a}).  These bright spots are the most
prominent elements of a network of filamentary emission, similarly to
what is observed, e.g., by
\cite{Dubner_et_al__1998__aj__NeutralHydrogenintheDirectionoftheVELASupernovaRemnant}.
Comparing the emissivity maps at $t = 480 \, \Jahr$ and $t = 900 \,
\Jahr$, we find that the UV and the X-ray emission start to fade after
the shock has ceased to heat the surfaces of the clouds; in particular
the inter-cloud medium is emitting only very weakly in the UV at $t =
900 \, \Jahr$.  A tentative comparison of the luminosities obtained in
the simulations with those estimated on current observations shows
that the emission of our models is roughly compatible with the
observed values, though it might be on the high end of the
observationally allowed range.  Note, that we do not consider any
influence of the foreground absorption on the luminosities derived by
simulations.

\begin{figure}
  \centering
  \includegraphics[width=9cm]{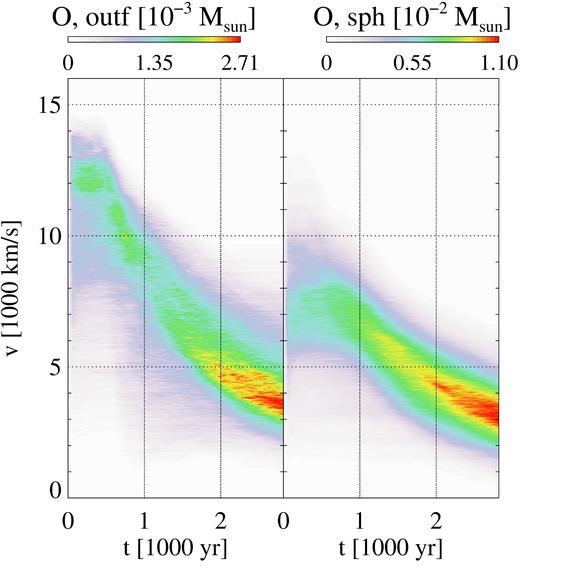}
  \includegraphics[width=9cm]{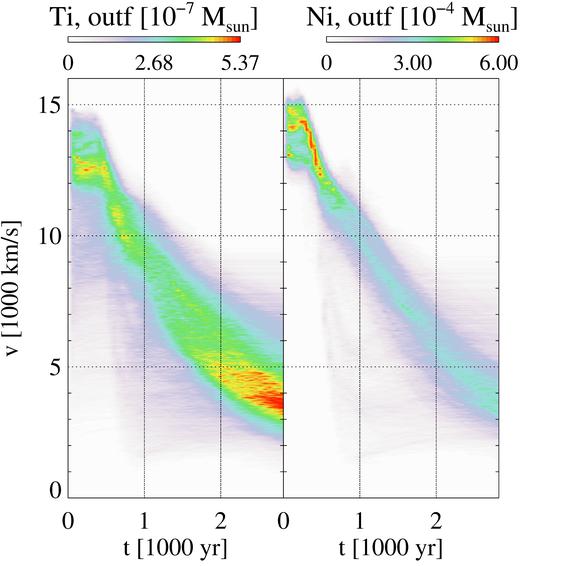}
  \caption{Velocity spectra of different elements in model
    \model{S25A}; \subpanel{top}: ${}^{16}\mathrm{O}$, outflow (left)
    and spherical (right) population; \subpanel{bottom}:
    ${}^{44}\mathrm{Ti}$ and ${}^{56}\mathrm{Ni}$ .  }
  \label{Fig:S25A-Histo}
\end{figure}

The complex flow patterns developing due to the interaction between
the shock and the clouds facilitate mixing of elements within the
ejecta.  This is reflected in both the spatial distribution of the
elements given by the tracer particles, and in the velocity spectra of
different elements.  We show the evolution of the velocity spectrum of
the oxygen ejected in the explosion in \figref{Fig:S25A-Histo}.  The
total oxygen mass amounts to $M_{\mathrm{O}} = 2.7\, \msun$, of which
$\sim 0.62\, \msun$ are in the outflow component.  During early epochs
the latter material has a significantly higher velocity ($v \approx
\zehnh{12}{3} \, \mathrm{km/s}$) than the oxygen in the spherical
component ($v \approx \zehnh{7}{3} \, \mathrm{km/s}$).  During the
expansion of the blast wave, and most notably when the explosion
interacts with the clouds, the ejecta are gradually slowing down.
This affects the outflow component more than the spherical component.
By $t \sim 1400 \, \Jahr$, the velocity difference between the two
components has reduced from an initial factor of $\sim 2$ to a factor
slightly above unity (cf.~the two distributions in the \subpanel{left
  panels} of \figref{Fig:S25A-Histo}).  As the particle distributions
in the \subpanel{top panels} of \figref{Fig:S25A-2d-900a} show, the
spatial distribution of particles belonging to the outflow and the
spherical component have similar features.  Both populations are
characterised by fingers protruding between the heavier clouds.  The
evolution of the distributions of nickel and titanium, which we place
in the initial model only in the outflow component, is very similar to
that of the outflow component of oxygen.  We note that the braking and
broadening of the Ti distribution is more pronounced than that of the
Ni distribution.  The fluid elements containing Ni have the highest
velocities, i.e., they are initially at the front of the outflow.
After the generation of fingers in the ejecta material, these fluid
elements remain at the largest radii and highest velocities.  The
mixing is most pronounced for the elements at intermediate radii.

\subsection{Variation of the model parameters}
\label{sSek:Vari}

\subsubsection{Explosion properties}

Below we assess the dependence of our results on the properties of the
SN explosion, i.e., ejecta mass, explosion energy, and the geometry of
the explosion.

\paragraph{Higher explosion energy}

\begin{figure}
  \centering
  \includegraphics[width=7.4cm]{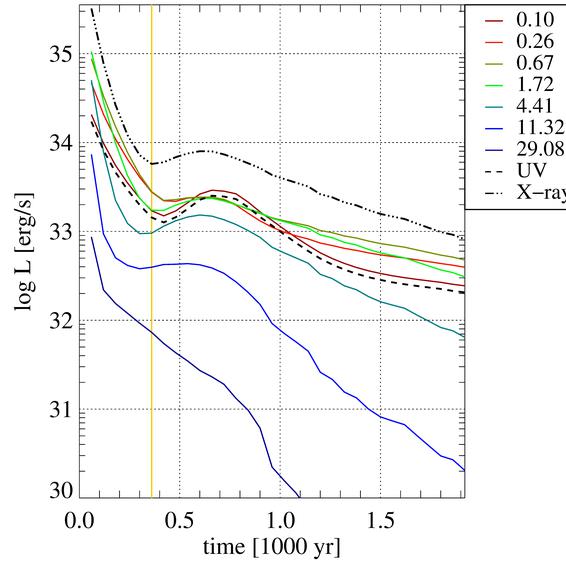}
  \includegraphics[width=7.4cm]{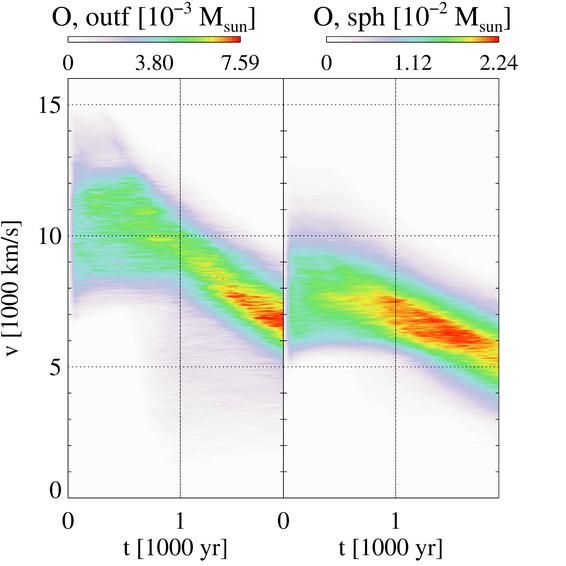}
  \includegraphics[width=7.4cm]{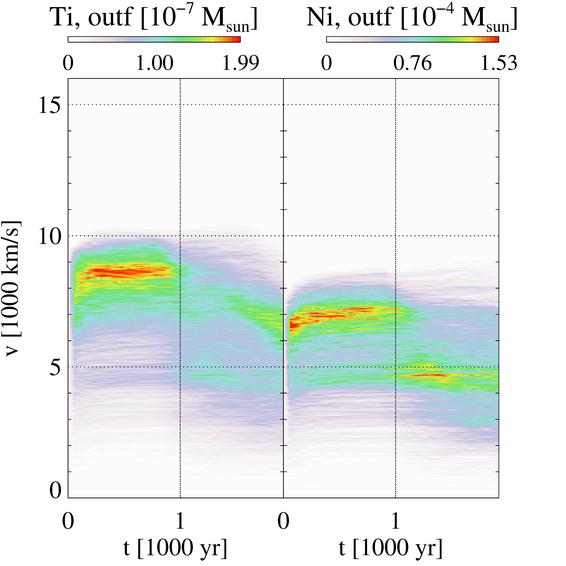}
  \caption{%%
    The light curve (\subpanel{top panel}) and velocity spectra
    (\subpanel{bottom two panels}) of model \model{S40} with both a
    higher explosion mass and explosion energy.
  }
  \label{Fig:S40A-LC}
\end{figure}

We compare the results described above to a simulation using a model
based on a progenitor of $40\, \msun$ (model \model{S40}).  According
to the simulations of
\cite{Maeda_Nomoto__2003__apj__Bipolar_SN:Nucleosynthesis_and_Implications_for_Abundances_in_EMP_Stars},
the explosion of this star yields an ejecta mass of $M_{\mathrm{SN}} =
10.2\, \msun$ and an explosion energy of $E_{\mathrm{SN}} =
\zehnh{10.9}{51}\, \mathrm{erg}$.  For these parameters, the expansion
of the shock wave and the mixing of the tracer particles proceeds very
much like in the reference model, hitting the large four clouds at the
same time.  This similarity can be attributed to the fact that the
ratio between explosion energy and mass is the same in both models.
The light curve of this model (\figref{Fig:S40A-LC})is qualitatively
similar to that of the reference model, but quantitative differences
exist.  We can identify the same phases of a very rapid decline from
the initial level, followed by a secondary rebrightening when the
shock wave interacts with the clouds, and a more gradual fading at
late times.  Throughout all stages, the total luminosity is higher
than in the reference model by a factor of order $\sim 2$, but the
hardness ratio between the X-ray and the UV emission is similar, with
the X-ray exceeding the UV by far.

\paragraph{Less energetic explosions}

\begin{figure}
  \centering
  \includegraphics[width=7.4cm]{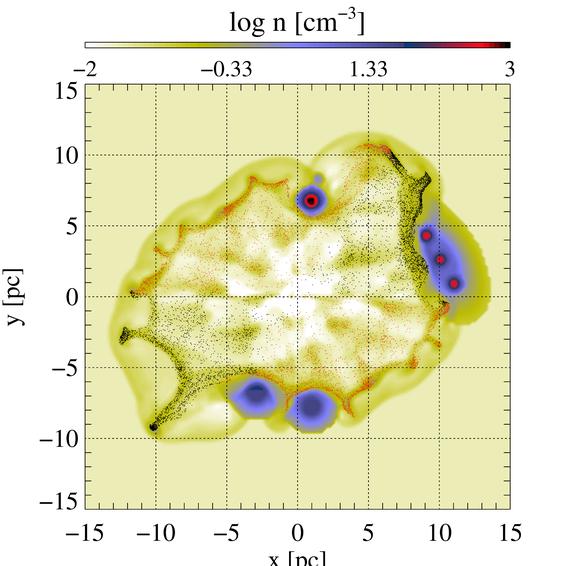}
  \includegraphics[width=7.4cm]{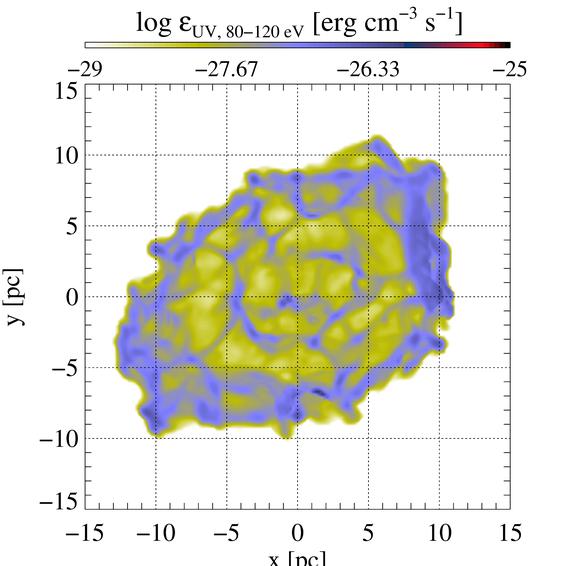}
  \includegraphics[width=7.4cm]{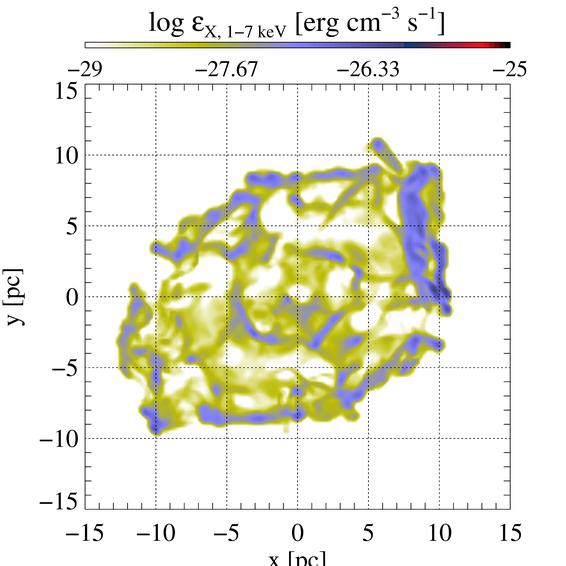}
  \caption{%%
    Same as \figref{Fig:S25A-2d-900a}, but for model \model{S25B}
    at time $t = 2400 \, \Jahr$.
  }
  \label{Fig:S25B-2d-3000a}
\end{figure}

\begin{figure}
  \centering
  \includegraphics[angle=0,width=7.4cm]{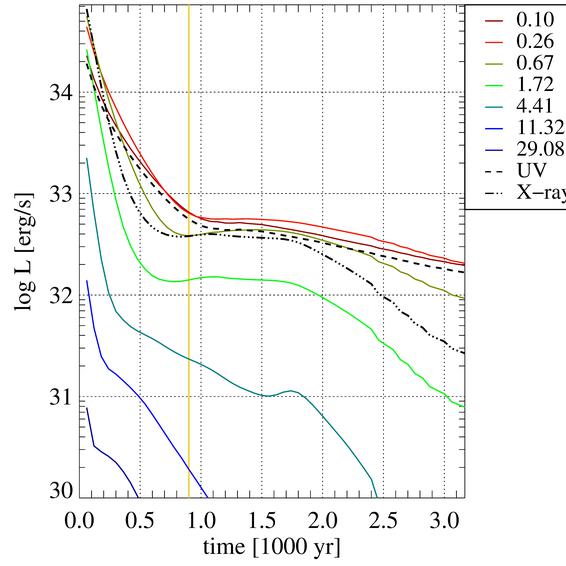}
  \includegraphics[angle=0,width=7.4cm]{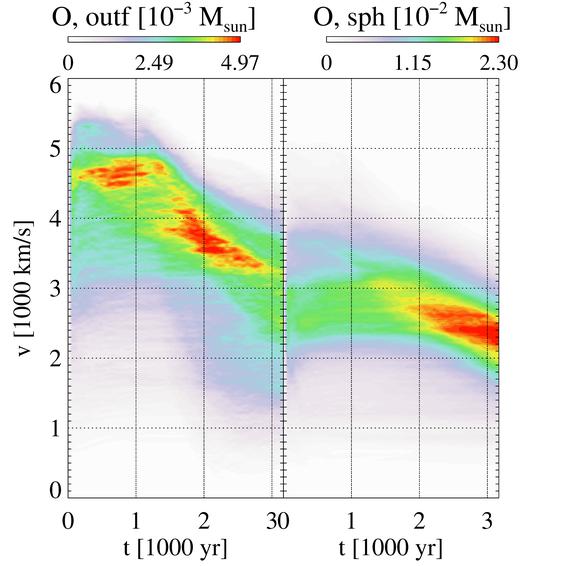}
  \includegraphics[angle=0,width=7.4cm]{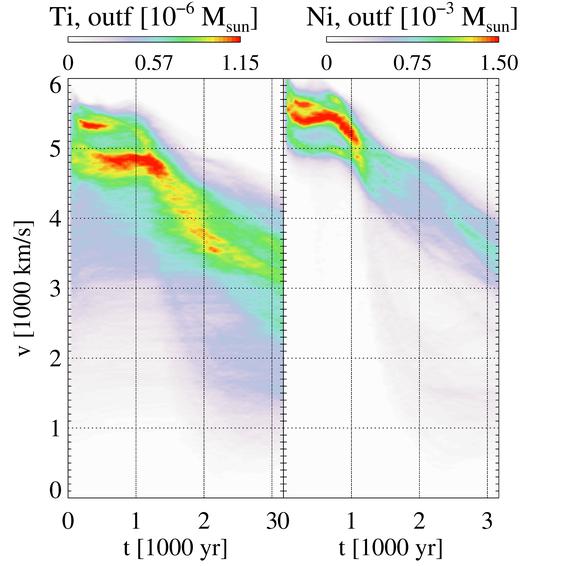}
  \caption{%%
    \subpanel{Top panel}: light curves of model \model{S25B}.  The solid line gives the
    total emission, and the dashed and dotted lines the emission in
    the X-ray and UV bands, respectively.  \subpanel{Middle and
        bottom panels}: velocity spectra for this model.
    }
  \label{Fig:S25B-LC}
\end{figure}

As we have seen, a higher explosion mass does not change the results
very much if the explosion energy is increased accordingly to keep
their ratio constant.  Models \model{S25B} and \model{S25BB} ejecting
a mass of $M_{\mathrm{SN}} = 6.0\, \msun$ but with an energy of
$E_{\mathrm{SN}} = \zehnh{1.0}{51} \, \mathrm{erg}$ and
$E_{\mathrm{SN}} = \zehnh{0.6}{51} \, \mathrm{erg}$, respectively,
only exhibits a qualitatively similar evolution, although at a slower
pace.  In model \model{S25B}, the blast wave hits the four large
clouds at a time of $t \sim 1500 \, \Jahr$ (first interactions with
clouds \CN{SE} and \CN{N} happen already at $t \sim \, 1000 \,
\Jahr$), i.e., about $900 \, \Jahr$ after the reference model with
higher explosion energy per unit mass.  This would correspond to an
explosion date around the $5^{\mathrm{nd}}$ century.  These events are
delayed by another $\sim 400 \, \Jahr$ when we decrease the explosion
energy in model \model{S25BB}.  While the dynamics is similar, the
light curves of the model (\figref{Fig:S25B-LC}) shows interesting
differences to the reference model.  The luminosity is, in general,
lower and softer, i.e., stronger in the UV than in the X-ray band.
The interaction between shock and clouds does not increase the
emission.  Instead, it gives rise to a long plateau phase of constant
emission after $t \sim 1000 \, \Jahr$, during which the UV and X-ray
luminosities are around $\zehnh{3}{32} \, \erg / \sek$.  Thus, the
high total luminosity and in particular the hardness of the emission
appears to be a feature distinguishing particularly strong explosions
from SNe of a lower energy.

A consequence of the lower explosion energy is the lower expansion
velocity of the fluid elements (see \figref{Fig:S25B-LC}), which is
reflected in the velocity spectrum of the tracer particles.
Furthermore, we find less mixing of particles from the outflow
population into the spherical one, mainly because the spherical
population of particles remains more concentrated in the central parts
of the SNR, with less particles escaping between the clouds close to
the centre of the explosion.

\begin{figure}
  \centering
  \includegraphics[width=7.8cm]{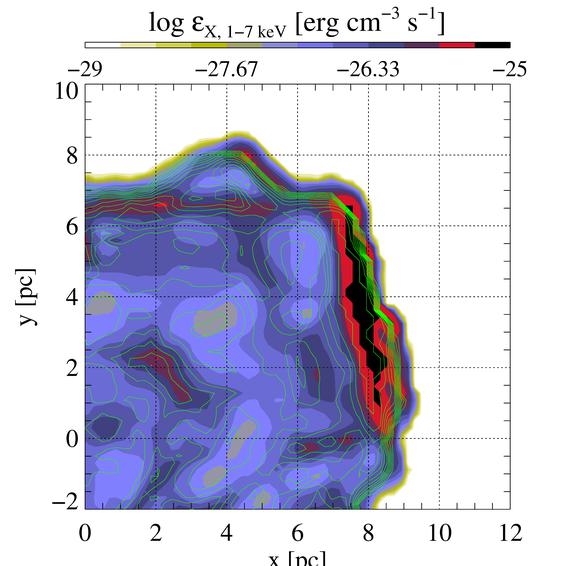}
  \includegraphics[width=7.8cm]{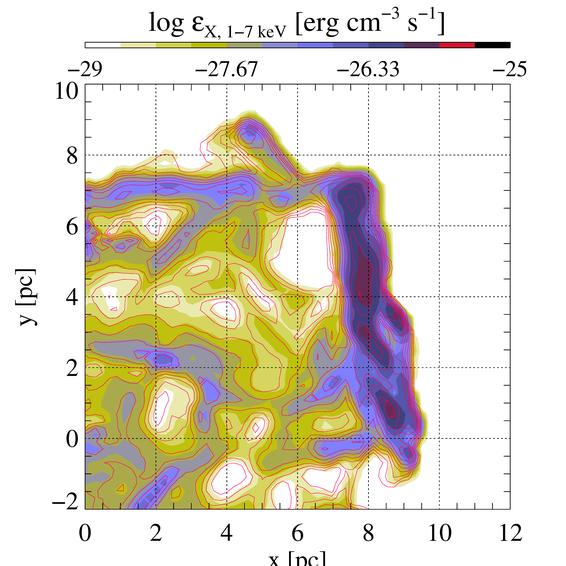}
  \includegraphics[width=5.5cm]{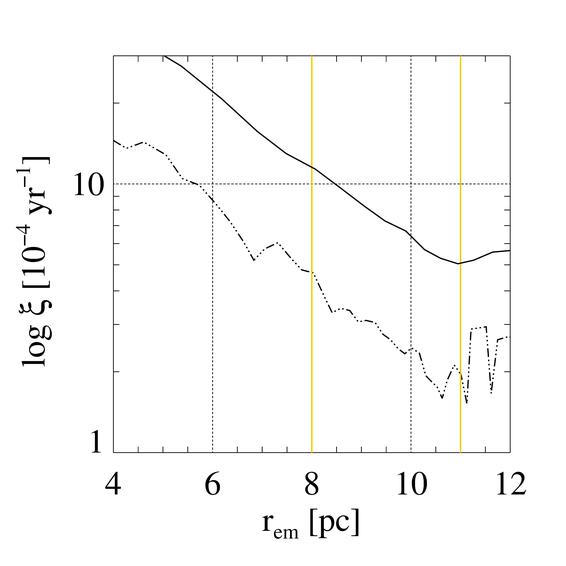}
  \caption{%%
    \subpanel{Top and middle panels}: the X-ray emissivity at the NW
    rim of the SNR in models \model{S25A} and \model{S25B}
    (\subpanel{top} and \subpanel{bottom} panels, respectively.  We
    show the emission at $t = 660 \, \Jahr$ (\model{S25A}) and $t =
    1860 \, \Jahr$ (\model{S25B}) in colours and the emission 60 years
    earlier (i.e., at $t = 600 \, \Jahr$ and $t = 1800 \, \Jahr$,
    respectively) with green (\subpanel{top}) and red
    (\subpanel{middle panel}) contours, respectively.  Colour and
    contours are using 12 logarithmically spaced levels between
    $\zehn{-29}$ and $\zehn{-25} \, \erg / \ccm \, \mathrm{s}$.
    \subpanel{Bottom panel}: the expansion rate of the emission front
    into the cloud \CN{NW} as a function of distance from the
    explosion for models \model{S25A} (solid) and \model{S25B}
    (dash-dot-dot-dotted line).  The vertical lines mark the innermost
    and outermost radii of the cloud.
  }
  \label{Fig:S25A-detail}
\end{figure}

We compare the X-ray emission at the outer rim at the time when the
shock hits the \CN{NW} cloud of this model to the reference model in
\figref{Fig:S25A-detail}.  The overall luminosity is higher for the
model with the stronger explosion, but the most striking difference is
the lower amount of substructure for lower explosion energies.  All
other parameters equal, a higher explosion energy leads to more
efficient heating of the gas and thus more emission, not only from the
surfaces of the clouds, but also from the tenuous gas in the interior
regions.  Besides the differences in appearance, higher explosion
energies also cause a larger shock velocity.  To reach the \CN{NW}
cloud at the stages shown in \figref{Fig:S25A-detail}, the
high-energy-explosion model requires about 600 years, while the shock
wave of the lower-explosion-energy one takes $\sim 1800$ years.  We
estimate the relative expansion rates, $\xi$, as the logarithmic time
derivative of the radius of the X-ray front along the X-ray-bright
ridge around $x \approx 9 \, \mathrm{pc}$, finding $\xi_{\model{S25A}}
\sim \zehnh{6...8}{-4} \, \Jahr^{-1}$ and $\xi_{\model{S25B}} \sim
\zehnh{2.5}{-4} \, \Jahr^{-1}$ for high and low explosion energies,
respectively.  For model \model{S25BB}, the value is the same within
the measuring uncertainties.

In the bottom panel of \figref{Fig:S25A-detail}, we compare the
estimated rate at which the emission front expands into the direction
of the cloud \CN{NW} as a function of its distance from the centre of
the explosion.  For each time step, we detected the radius,
$r_{\mathrm{em}}$ of the emission front and then computed the
expansion rate, $\xi = \partial_{t} \log r_{\mathrm{em}}$.  During the
entire evolution, the emission front slows down for bot models.
During the interaction of the shock wave with cloud \CN{NW},
\model{S25A} has a much higher expansion rate than \model{S25B}.  In
both models, the rate at which the shock wave expands into cloud
\CN{NW} is lower by about a third w.r.t.~free expansion in a direction
without any clouds.

Comparing these values to $\xi \sim \zehnh{2.3}{-4} \, \Jahr^{-1}$
derived by
\cite{Katsuda__2008__apjl__TheSlowX-RayExpansionoftheNorthwesternRimoftheSupernovaRemnantRXJ0852.0-4622},
our simulations appear to favour an explosion energy closer to the
canonical $\zehn{51} \, \erg$ than to hypernova-like energies.  We
point out that this is only a shock velocity in the cloud, but not an
estimate of the expansion rate of the SNR shock wave, which can be
much higher.

Because taking into account just thermal bremsstrahlung, our analysis
is limited to a part of the radiation emitted by the SNR only.
Although a full treatment including all radiation processes is beyond
the scope of this work, we can estimate the impact of our restriction
to thermal bremsstrahlung on our results. To this end we employ a
criterion for the most important missing process, non-thermal
synchrotron radiation, given by
\cite{Vink__2012__aapr__Supernovaremnants:theX-rayperspective}.  X-ray
synchrotron emission requires a shock velocity exceeding 2000\,km/s.
This condition is fulfilled for the propagation of the shock in the
inter-cloud regions, but is violated for the shock of model
\model{S25B} hitting cloud \CN{NW}, where $v_{\mathrm{shock}} \sim
1500 \, \km / \sec$. In contrast, in \model{S25A} $v_{\mathrm{shock}}
\sim 4000$ in that cloud exhibiting considerable fluctuations in time
and space.  Consequently, our models would emit appreciable amounts of
synchrotron radiation in the \CN{NW} cloud only for a high explosion
energy.  Hence, the detection of non-thermal emission from that region
is consistent with a high-energy supernova, albeit no conclusive
evidence for it.  Going beyond the qualitative statement that our
models should produce X-ray synchrotron emission is difficult.
Without the ability to produce detailed synchrotron maps from our
numerical simulations, we can only speculate about the spatial
distribution of the emission.  We might expect a qualitative
similarity between the synchrotron emissivity and the thermal
emissivity maps presented in \figref{Fig:S25A-detail}, i.e., extended
filamentary structures behind the shock.  However, we can only resolve
this issue with a more elaborate modelling approach.  Most properties
of synchrotron radiation depend on the magnetic field in the emitting
gas.  Since our simulations do not include a magnetic field, we can
only rely on estimates derived from observations, e.g., by
\cite{Kishishita_et_al__2013__aap__NonthermalemissionpropertiesofthenorthwesternrimofsupernovaremnantRXJ0852.0-4622}.
However, the properties of the synchrotron radiation beyond the
field-independent cut-off energy are determined by additional
parameters which we cannot extract from our simulations.  Therefore,
we do not attempt a more accurate analysis here.

\paragraph{Spherical explosions}
To explore the effects of the geometry of the explosion, we simulated
models \model{S25As} and \model{S25Bs}, explosions with a mass of
$M_{\mathrm{SN}} = 6.0 \, \msun$ and energies of $E_{\mathrm{SN}} =
6.7$ and $\zehnh{1.0}{51} \, \erg$, but exploding spherically.  In the
reference model, we deliberately point the explosion spheroid towards
cloud \CN{NW}.  The combination of this orientation and the axis ratio
of the explosion yields a roughly equal time at which the blast wave
collides with all four major clouds.  In the spherical case, on the
other hand, the different distances of these clouds from the explosion
center lead to larger differences in the times of impact: for high
explosion energy, the clouds \CN{S}, \CN{SE}, and \CN{N} are hit at $t
\approx 500 \, \Jahr$, wheras it takes another $300 \, \Jahr$ for the
shock to arrive at cloud \CN{NW}.  In the case of low explosion
energies, these events take place in the same order, but roughly $2000
\, \Jahr$ later.

Of course, the most obvious observation signature of a spherical
explosion would be a spherical appearance of the SNR.  However, in an
inhomogeneous ISM, the shape of the SNR may become distorted during
its expansion.  Furthermore, as we have seen above, even a bipolar
explosion can lead to a moderate axis ratio of the remnant after
several hundreds of years.  Additionally, parts of an SNR might not be
easily accessible observationally, rendering a clear identification of
the shape difficult.  Thus, additional observational information might
be helpful to obtain the properties of the SN that created the
remnant.  We consider in the following the most distinct properties of
the light curves of these explosions.

\begin{figure*}
  \centering
  \includegraphics[angle=0,width=7.4cm]{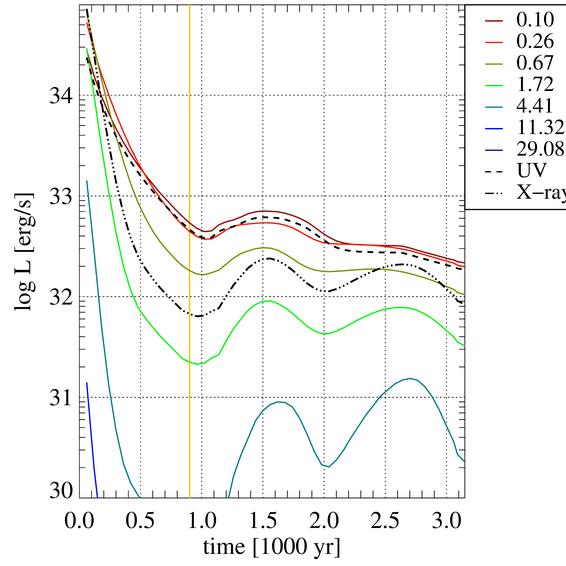}
  \includegraphics[angle=0,width=7.4cm]{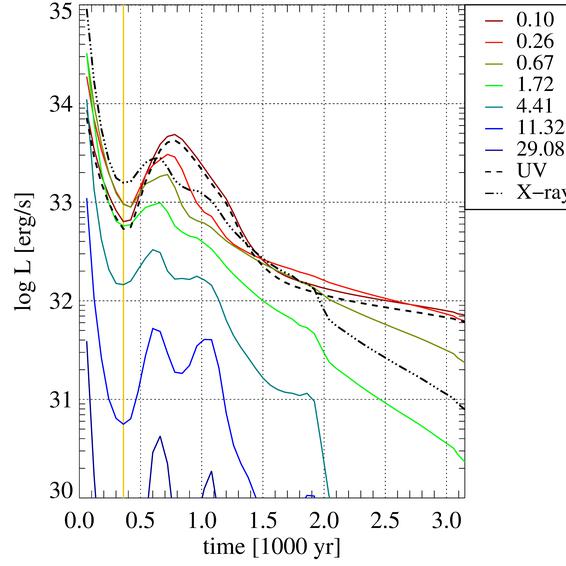}
  \includegraphics[angle=0,width=7.4cm]{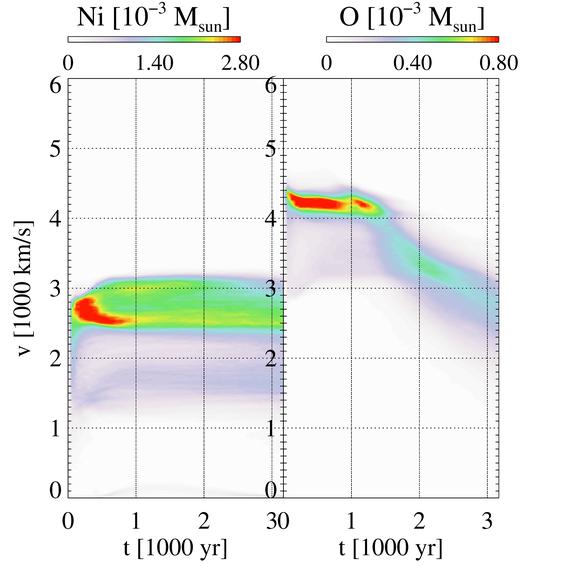}
  \includegraphics[angle=0,width=7.4cm]{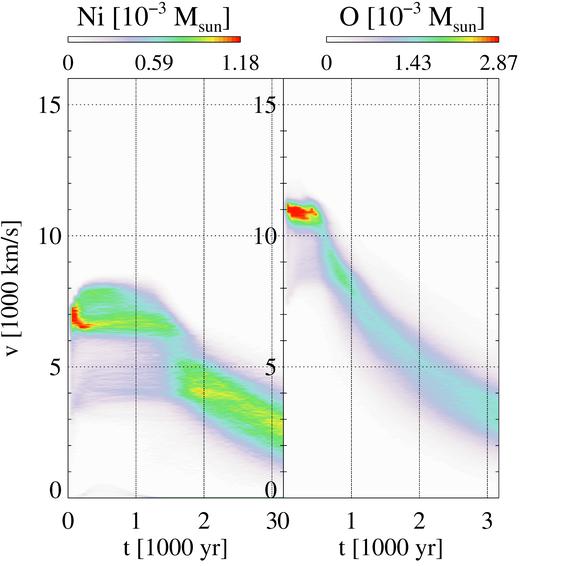}
  \caption{%%
    The light curves (\subpanel{top}) and velocity spectra of nickel
    and oxygen (\subpanel{bottom}) of spherical models
    \model{S25Bs} and \model{S25As} with an explosion energy of
    $E_{\mathrm{SN}} = \zehnh{1.0}{51} \, \erg$ (\subpanel{left
      panels}) and $E_{\mathrm{SN}} = \zehnh{6.7}{51} \, \erg$
    (\subpanel{right panels}).
  }
  \label{Fig:S25As-LC}
\end{figure*}

We display the light curves of models \model{S25Bs} and \model{S25As}
in \figref{Fig:S25As-LC}.  Since a large fraction of the total
luminosity is emitted by the large clouds, the differences in the
dynamics w.r.t.~the anisotropic models manifest themselves in
modifications of the photon emission.  We observe again the common
phases of a fast decline early on, a plateau or rebrightening during
the interaction with the clouds, and a fading afterwards.  Similarly
to models \model{S25A} and \model{S25B} discussed above, the spectra
are harder for higher explosion energies, though even for the
high-energy explosion the emission is not as strongly dominated by the
X-ray emission as for the anisotropic explosions.  Note also that the
high-energy model \model{S25As} emits stronger in UV than in X-rays
between $t \approx 600 \, \Jahr$ and $t \approx 900 \, \Jahr$, i.e.,
when the shock hits the clouds.  The low-energy model \model{S25Bs}
has a very soft emission at late times, since the X-ray luminosity
decreases rather rapidly after $t \sim 2500 \, \Jahr$.

\subsubsection{Properties of ISM and clouds}

\paragraph{Clouds}

\begin{figure*}
  \centering
  \includegraphics[width=7.4cm]{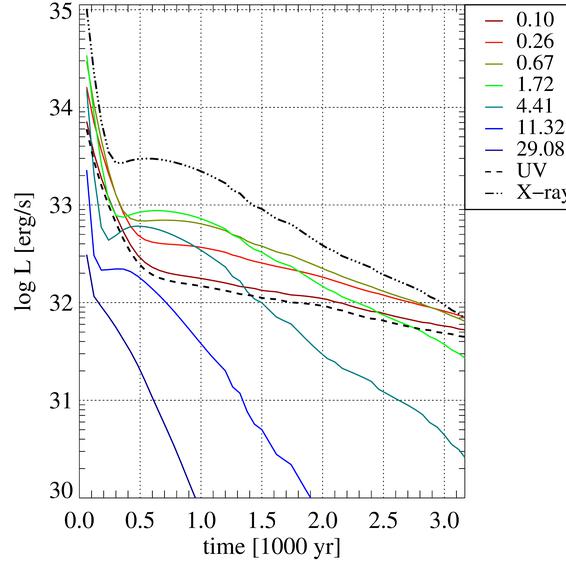}
  \includegraphics[width=7.4cm]{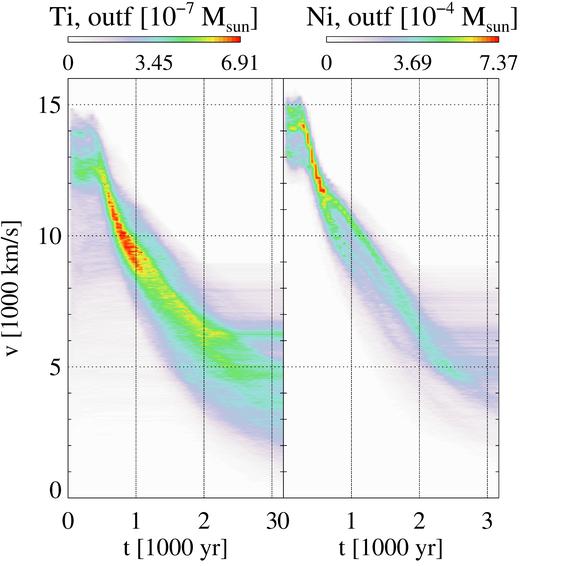}
  \includegraphics[width=7.4cm]{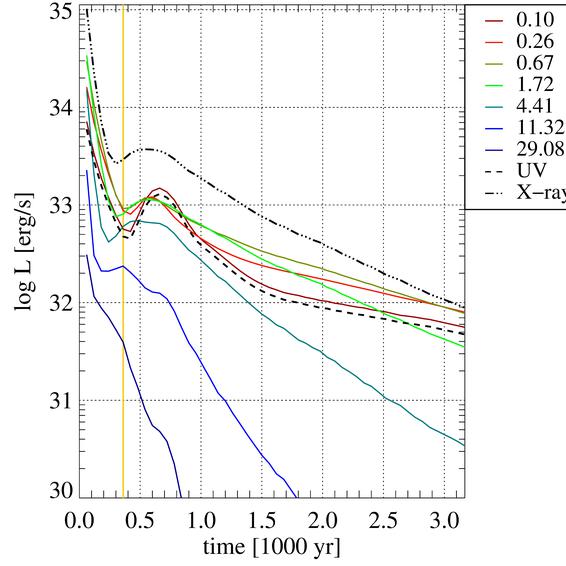}
  \includegraphics[width=7.4cm]{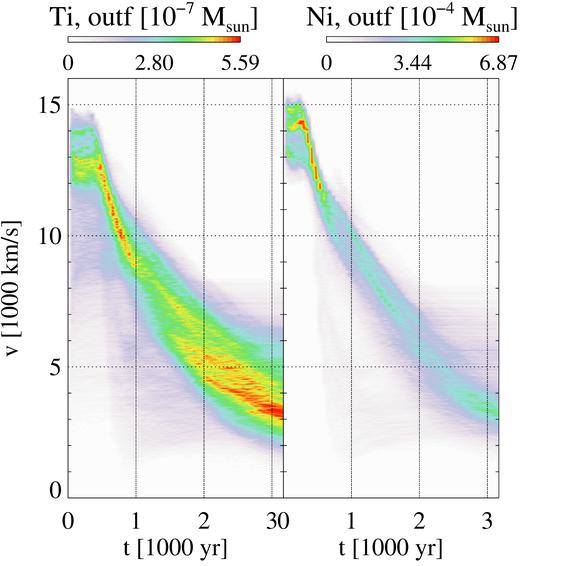}
  \includegraphics[width=7.4cm]{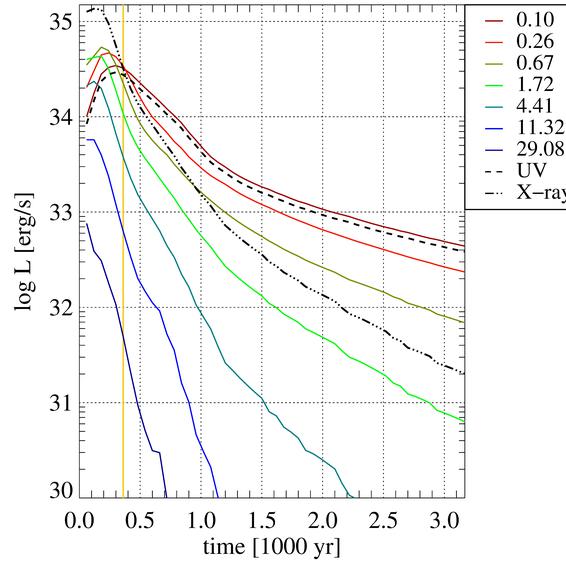}
  \includegraphics[width=7.4cm]{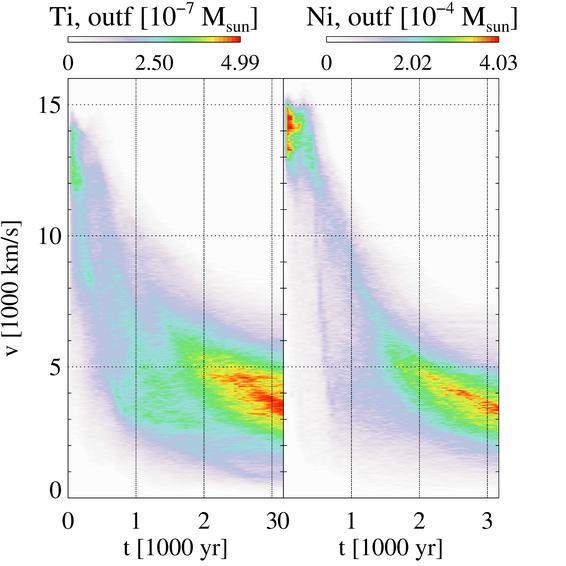}
  \caption{%%
    \subpanel{Left panels}: the light curves of models \model{S25Anc}
    (\subpanel{top}), \model{S25A4c} (\subpanel{middle}), and
    \model{S25AC} (\subpanel{bottom}), respectively.  \subpanel{Right
      panels}: the velocity spectra of Ti and Ni of the same models.
  }
  \label{Fig:S25Av-LC}
\end{figure*}

The position and the distribution of sizes and densities of small
clouds is poorly constrained from observations. Therefore, we can only
hope to model the real state of the ISM in a very approximate manner
by inserting clouds of random sizes at random points into the ISM.  In
two models with $M_{\mathrm{SN}} = \zehnh{6}{51} \, \msun$ and
$E_{\mathrm{SN}} = \zehnh{6.7}{51} \, \erg$, we did not include any
clouds at all (\model{S25Anc}) or only the four large ones for which we
have observational indications (\model{S25A4c}).  Consequently, there
is no mixing until $t \approx 600 \, \Jahr$ when the shock wave hits
cloud \CN{N}.

In another series of models (models \model{S25Anc}, \model{S25A4c},
\model{S25AC}, \model{S25Aw}, \model{S25Ad0.1}, \model{S25Ad1.0}, and
\model{S25Ad10}), rather than reducing the number of clouds, we
explored the consequences of the presence of more and denser clouds in
the vicinity of the explosion, while keeping the explosion energy and
mass the same as in the reference model \model{S25A}.

Comparing the results of these models among each other and to those of
the model \model{S25A}, we can conclude that the expansion velocity of
the blast wave is not affected strongly by the absence of a multitude
of small clouds.  The post-shock flow, on the other hand, shows much
less features without the random population of clouds.  This is
reflected in the distribution of the particles tracing the mixing of
elements in velocity space.  We find much less dispersion of both the
spherical and the outflow components without the additional clouds.
At $t \approx 500\, \Jahr$, e.g., the distributions of titanium and
nickel in velocity space exhibit a rather pronounced maximum at $v
\approx 10^4\, \mathrm{km/s}$.  A certain broadening of the
distributions occurs after the interaction with the four clouds took
place. Afterwards, the distribution retains its shape until the end of
simulations at $t = 3000\, \Jahr$.  The impact of the ejecta onto
clouds clouds leads to both up-scattering and down-scattering of
particles in the velocity space. While some parts of the flow are
decelerated in interactions with the clouds, other parts are
accelerated when the flow is squeezed through the less dense, narrow
channels existing between the clouds. On average, though, the energy
loss exceeds the energy gain, and the total kinetic energy is
decreasing by about 10 \% when the random clouds are included.

In model \model{S25AC}, we placed 1200 small clouds with a density $n
= 20 \, \iccm$ randomly in the simulation domain.  Though this is most
likely more than can be expected in the vicinity of the actual SNR, we
used this model to explore the general effect of many small clouds.
Most prominently, the expansion of the explosion is slowed down by the
large number of interactions the shock wave suffers.  The luminosity
is very high during the entire evolution.  Furthermore, we find rather
soft spectra, the UV band exceeding the X-ray emission after $t
\approx 400 \, \Jahr$ and decaying slower than in the latter band (see
\figref{Fig:S25Av-LC}, \subpanel{bottom left panel}).  The frequent
interactions between shock and clouds broaden the velocity spectra of
Ti and Ni considerably, and they lead to rather slow ejecta (see
\figref{Fig:S25Av-LC}, \subpanel{bottom left panel}).

We find a clear correlation between the photon emission and the cloud
population (see \figref{Fig:S25Av-LC}, \subpanel{left panels}).
Without any clouds (\subpanel{top panel}), the emission has a high
hardness ratio between X-ray and UV.  The most striking difference to
the light curves of other models is probably the very slow decline of
the UV emission after $t \sim 500 \, \Jahr$.  The X-ray luminosity, on
the other hand, decreases exponentially during this time.

\paragraph{Varying the density and temperature of the ISM}
Since the density and temperature of the ISM surrounding the ejecta
and the clouds are not well constrained by observations, we have
varied these two parameters in our simulations as well.  While the
temperature of the ISM appears to have only a minor influence on the
dynamics of the gas, increasing the gas density leads to remarkable
differences from the reference model.  In model \model{S25Aw}, we
start the explosion in a medium of a temperature of $T_{\mathrm{ISM}}
= 1000\, \mathrm{K}$ instead of the standard value of
$T_{\mathrm{ISM}} = 10\, \mathrm{K}$.  This choice does neither affect
the time at which the SN shock wave hits the four main clouds, nor the
emission from the gas, nor the velocity spectra of the different
elements significantly.  Thus, though we assumed a very cold ISM in
all our simulations, our results are valid as well for a warm ISM.

\begin{figure}
  \centering
  \includegraphics[width=7.4cm]{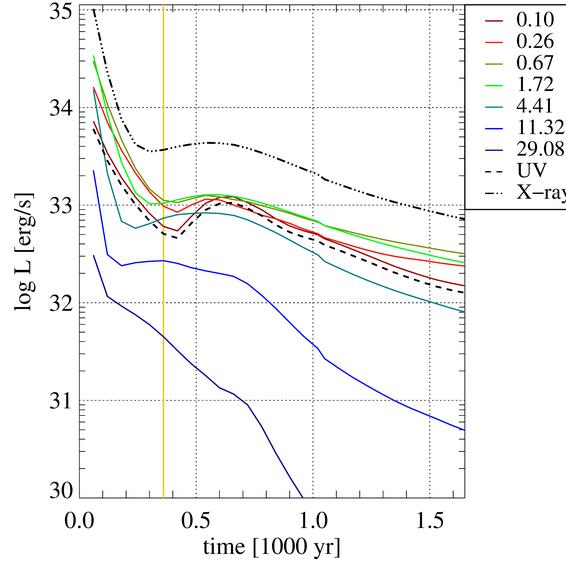}
  \includegraphics[width=7.4cm]{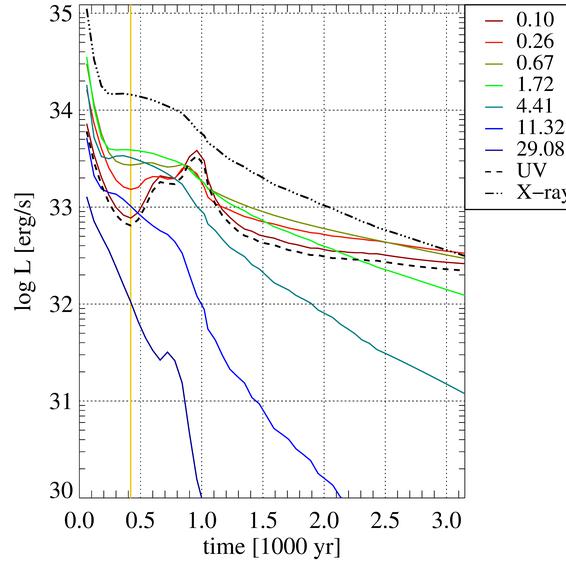}
  \includegraphics[width=7.4cm]{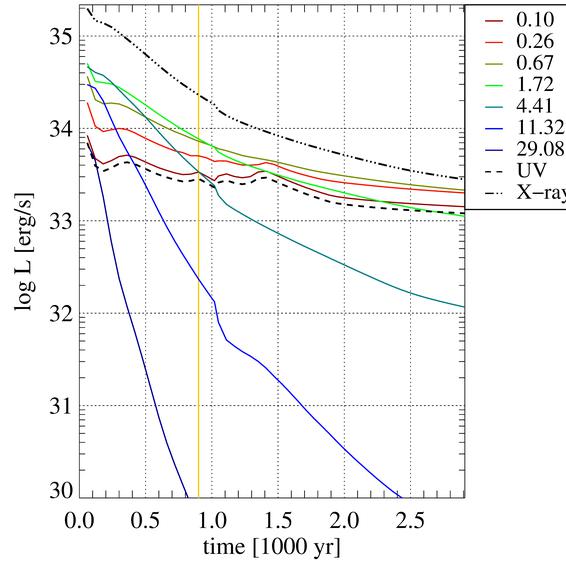}
  \caption{%%
    The light curve of models \model{S25Aw} (warm ISM;
      \subpanel{top}), \model{S25Ad0.1} ($n_{\mathrm{ISM}} = 0.1 \,
    \iccm$; \subpanel{middle}),  \model{S25Ad1.0} ($n_{\mathrm{ISM}}
    = 1.0 \, \iccm$; \subpanel{bottom}).
  }
  \label{Fig:S25d-LC}
\end{figure}

A denser ISM, on the other hand, delays the expansion of the shock
wave.  We have simulated a model (\model{S25Ad0.1} with a moderately
increased ISM density of $n_{\mathrm{ISM}} = 0.1\, \iccm$ rather than
$n_{\mathrm{ISM}} = 0.025\, \iccm$.  While the hydrodynamics of the
expansion and the mixing are hardly affected by this change, we find a
stronger emission in all bands (see \subpanel{middle panel} of
\figref{Fig:S25d-LC}).

A further increase of the ambient density (models \model{S25A-d1.0}
and \model{S25A-d10} with $d = 1 \, \iccm$ and $d = 10 \, \iccm$,
respectively) leads to a modified dynamics.  The blast wave, sweeping
up ambient matter much faster in this model, is decelerated more
efficiently and reaches the four main clouds at a much later time:
cloud \CN{NW} at $t \approx 1300\, \Jahr$ for model \model{S25Ad1.0}
and $t = 3000 \, \Jahr$ for model \model{S25Ad1.0}, respectively.
Although the evolution is considerably slower than in the reference
model, the mixing is similar to that of the reference model.  We find
that the tendency of stronger emission for increasing ISM density
continues here: the luminosity of model \model{S25A-d1.0} at the time
of contact with cloud \CN{NW} is $L_{\mathrm{X}} \sim \zehn{34} \,
\erg / \mathrm{s}$, and the UV band shows a flat light curve exceeding
$L_{\mathrm{UV}} \sim \zehn{32} \, \erg / \mathrm{s}$ during the
entire time of the simulation.  Because of the enhanced density in the
interior regions of the SNR, the luminosity is dominated far less by
the cloud than in the reference model, showing instead a comparably
featureless volume-filling emission.

\subsubsection{Distance of the SNR}
The distance to Vela Jr.~is a matter of intense debate, with estimates
between $\sim 300 \, \mathrm{pc}$ and over $1000 \, \mathrm{pc}$.
The models described above were performed assuming a low value for the
distance, $D = 300 \, \mathrm{pc}$.  Here, we show the results of a
simulation where we placed the reference model at a distance of $D =
750 \, \mathrm{pc}$ (model \model{S25AD}).

The distance enters our setup and the interpretation of our results
in two ways:
\begin{itemize}
\item Given the celestial coordinates of the clouds, their distance
  from the centre of our simulation grid are proportional to the
  assumed distance.  Thus, if we place an explosion at  a larger
  distance without changing its parameters like mass and energy, the
  time required until the blast wave interacts with the clouds,
  producing the current observational appearance, is considerably
  longer.
\item A larger distance means that the apparent brightness of the
  remnant in UV and X-rays corresponds to a higher absolute
  luminosity.
\end{itemize}

\begin{figure*}
  \centering
  \includegraphics[angle=0,width=8.0cm]{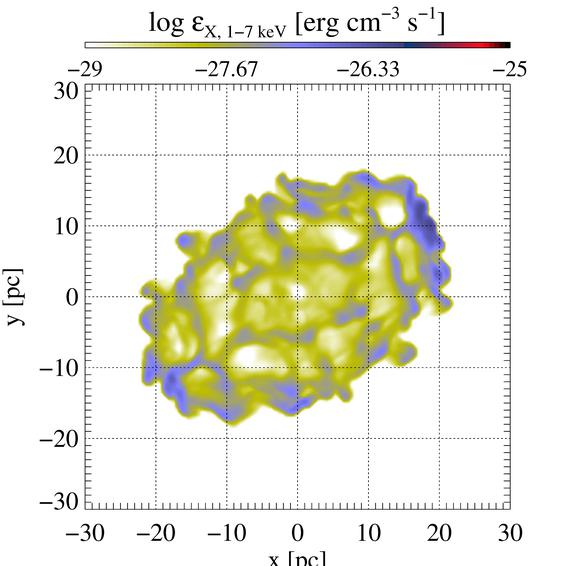}
  \includegraphics[angle=0,width=8.0cm]{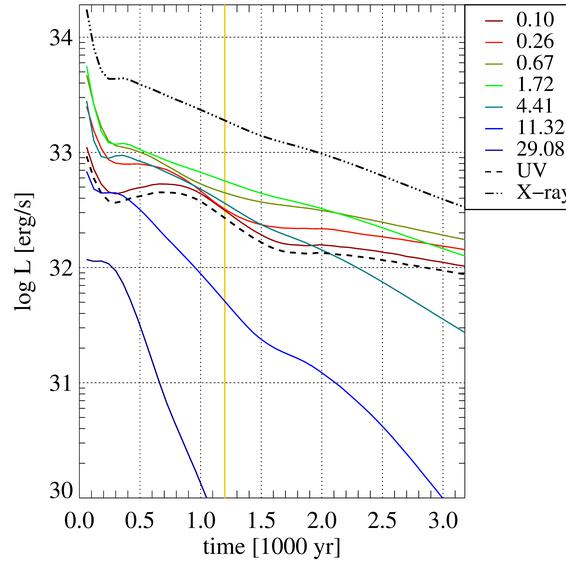}
  \includegraphics[angle=0,width=8.0cm]{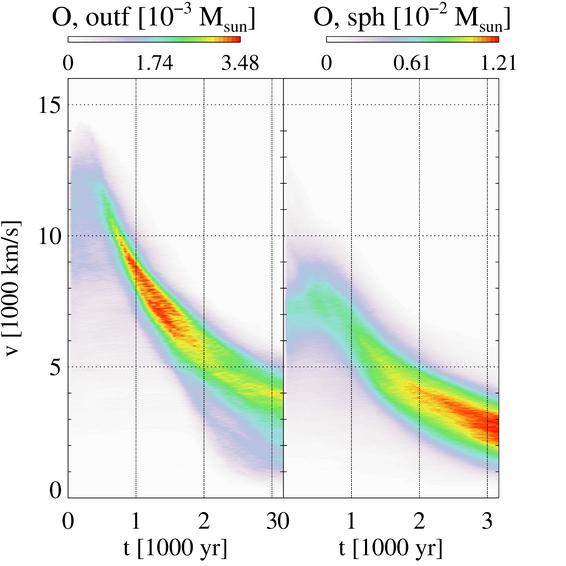}
  \includegraphics[angle=0,width=8.0cm]{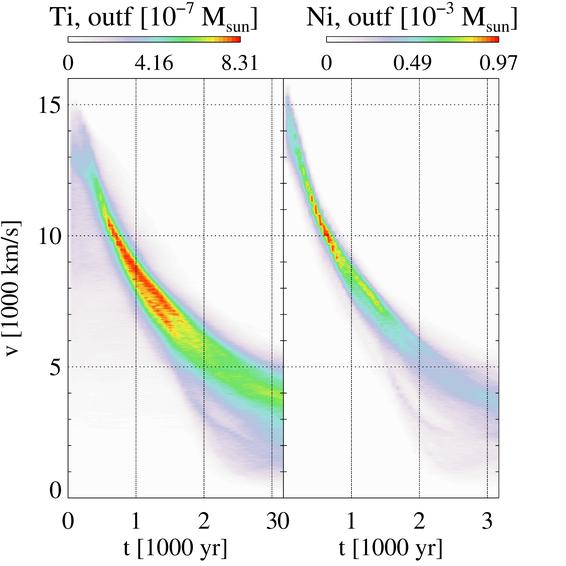}
  \caption{%%
    Model \model{S25AD} (SNR located at larger distance): X-ray
    emissivity map at $t = 2220 \, \Jahr$ (\subpanel{top left}), light
    curves (\subpanel{top right}), and velocity spectra of oxygen
    (\subpanel{bottom left}: outflow; \subpanel{bottom right}:
    spherical components), nickel and titanium.
  }
  \label{Fig:S25A-distant}
\end{figure*}

The interaction with the clouds starts with cloud \CN{SE} at
$t \approx 1300 \, \Jahr$; it takes about 400 yr more to reach cloud
\CN{NW}.  For such a model, current appearance of Vela Jr. would be compatible
with the supernova explosion around the year 0 AD.  Overall this model shows the
same dynamics as the reference model, and therefore the observational
appearance at the time of the shock-cloud interaction is rather
similar to the reference model (see \figref{Fig:S25A-distant},
\subpanel{top left panel}).  However, after a much longer expansion, the
energy and mass of the explosion are distributed over a much larger
volume, and therefore, the absolute values of the emissivity is less
than for the same model assuming a lower distance.  Apart from this
scaling, the emission properties are comparable.  In particular, the
model with a high explosion energy, exhibits a rather hard
spectrum during the entire evolution we considered here (cf.~the UV
and X-ray lightcurves in the \subpanel{top right panel} of
\figref{Fig:S25A-distant}).  Additionally, as can be seen in the
\subpanel{bottom panels} of \figref{Fig:S25A-distant}, the velocity
spectra are very similar to our reference model, but the long time of
expansion before the shock wave encounters the clouds lead to a
somewhat less pronounced broadening of the population of fluid
elements in velocity space.

\section{Summary and conclusions}
\label{Sek:SumCon}

We have performed a series of hydrodynamic simulations of the
expansion of a supernova explosion into an anisotropic ambient
interstellar medium.  In choosing the properties of the ISM, we had
the clouds observed in the supernova remnant Vela Jr.\ in mind.  The
ISM in this SNR is dominated by four large clouds at roughly equal
(projected) distance from the centre of the remnant.  In addition to
these clouds, we distributed a number of smaller, less dense clouds
randomly in the vicinity of the explosion site.

The age of the SNR, and the mass and energy of the ejecta are only
poorly constrained by observations.  The goal of our simulations was
to explore the parameter space of these properties of the explosion
and to compare the simulation results to the UV and X-ray emission
observed in the SNR
\citep{Aschenbach__1998__nat__Discoveryofayoungnearbysupernovaremnant,
  Aschenbach_et_al__1999__aap__ConstraintsofagedistanceandprogenitorofthesupernovaremnantRXJ0852.0-4622_GROJ0852-4642,
  Tsunemi_et_al__2000__pasj__OverabundanceofCalciumintheYoungSNRRXJ0852-4622:EvidenceofOver-Productionof$44$Ti,
  Slane_et_al__2001__apj__RXJ0852.0-4622:AnotherNonthermalShell-TypeSupernovaRemnant(G266.2-1.2),
  Nichols_Slavin__2004__apj__ShockedCloudsintheVelaSupernovaRemnant,
  Iyudin_et_al__2005__aap__XMM-NewtonobservationsofthesupernovaremnantRXJ0852.0-4622_GROJ0852-4642,
  Iyudin_et_al__2007__ESASP__MultiwavelengthAppearanceofVelaJr.:IsituptoExpectations,
  Bamba_et_al__2005__apj__ChandraObservationsofGalacticSupernovaRemnantVelaJr.:ANewSampleofThinFilamentsEmittingSynchrotronX-Rays,
  Nishikida_et_al__2006__apjl__Far-UltravioletSpectralImagesoftheVelaSupernovaRemnant,
  Pannuti_et_al__2010__apj__Non-thermalX-rayEmissionfromtheNorthwesternRimoftheGalacticSupernovaRemnantG266.2-1.2(RXJ0852.0-4622),
  Iyudin_et_al__2010__aap__SearchforbroadabsorptionlinesinspectraofstarsinthefieldofsupernovaremnantRXJ0852.0-4622(VelaJr.),
  Pakhomov_et_al__2012__mnras__InterstellarabsorptionsandshockedcloudstowardsthesupernovaremnantRXJ0852.0-4622,
  Kim_et_al__2012__apj__Far-ultravioletSpectralImagesoftheVelaSupernovaRemnant:SupplementsandComparisonswithOtherWavelengthImages}.
Our main focus was on very energetic bipolar explosions that belong to
the class of hypernovae.  We initiated the explosion by placing the
ejecta into a small spheroid (of roughly 1\,pc size) at the centre of
the computational domain.  The explosion energy was mostly (95\%)
assumed to be in the form of kinetic energy.  We chose the properties
of the explosion (mass, energy, and axes ratio of the spheroid)
according to the simulations by
\cite{Maeda_Nomoto__2003__apj__Bipolar_SN:Nucleosynthesis_and_Implications_for_Abundances_in_EMP_Stars},
who computed the late phases of the hypernova explosions of main
sequence stars of 25 and 40\,$\msun$. These authors provide masses and
explosion energies as well as abundances of different elements in the
ejecta.

We based the comparison of our simulation results with the
observations on the following diagnostics:
\begin{enumerate}
\item The morphology of the ejecta and its interactions with the ISM
  is a direct result of the hydrodynamics of the explosion.  In
  particular, we focus on the time scales of the expansion of the
  blast wave and on the development of small-scale structures.
\item At $t = 0$, we place a large number of test particles in the
  ejecta and assign different chemical compositions to them based on
  their initial positions.  The particles are advected passively with
  the flow.  The positions of these particles at later times allow us
  to gain a detailed description of the mixing of different elements
  in the ejecta caused by the decay of the flow into smaller
  structures.  We construct time-dependent velocity spectra of oxygen,
  titanium and nickel that can be compared to observations.
\item We compute the thermal bremsstrahlung emission from the
  gas in different electromagnetic bands and construct mock light
  curves.  Since the collision time of electrons and ions can be long
  in the tenuous gas forming the interior of the SNR compared to its
  evolutionary time scales, we have to account for the non-equilibrium
  ionisation (NEI) of the gas.  Adding an equation for the energy of
  the electron gas to the system of hydrodynamic equations and
  modelling the coupling of the two species with a simplified ansatz
  for the collision time
  \citep{Spitzer__1962__PhysicsofFullyIonizedGases}, we compute the
  electron temperature, which is, in general, different from the
  temperature of the ions.  Based on these variables, we compute the
  photon emission due to bremsstrahlung in different bands at UV and
  X-ray energies
  \citep{Tucker_Koren__1971__apj__RadiationfromaHigh-TemperatureLow-DensityPlasma:theX-RaySpectrumoftheSolarCorona}.
  We also look for correlations between the model parameters and the
  light curves and spectral properties of the models.
\end{enumerate}

The main results of our simulations can be summarised as follows:
\begin{enumerate}
\item The supernova ejecta expanding into the ISM hits the four
  large clouds after a few hundred years.  For our reference model, a
  hypernova-like bipolar explosion with an ejecta mass of
  $M_{\mathrm{ex} } = 6.0\, \msun$ and an energy of $E_{\mathrm{ex}} =
  6.7 \times 10^{51}\, \mathrm{erg}$, this happens at roughly $t \sim
  600\, \Jahr$.  Current observations of the SNR Vela Jr.\ indicate
  that this moment of first interaction between the blast wave and the
  clouds has occurred already.  Though it is difficult to estimate the
  exact time that has passed since, we can conclude that the current
  appearance of the SNR would be consistent with a very energetic
  supernova explosion in the $12^{\mathrm{th}}$ century, with an
  uncertainty of a few hundred years.  The uncertainty of this result
  is rather large, as can be seen when varying the properties of the
  explosion and of the SNR environment.  A lower explosion energy closer to the
  canonical value of $10^{51}\,\mathrm{erg}$ or a higher density of
  the inter-cloud ISM delay the impact of the blast wave on the clouds
  significantly, and we would estimate a time of explosion about 2000
  years earlier, consistent with the speculations by
  \cite{Katsuda_et_al__2009__AdvancesinSpaceResearch__IsVelaJrayoungsupernovaremnant}.
  Besides the geometry of
  the SNR, we can estimate the rate at which the X-ray bright front
  marking the interaction of the blast wave and the clouds expands.
  The results of our models with a canonical explosion energy are in agreement with the value derived from observations by
  \cite{Katsuda__2008__apjl__TheSlowX-RayExpansionoftheNorthwesternRimoftheSupernovaRemnantRXJ0852.0-4622},
  while a higher explosion energy gives rise to a faster expansion.
\item The emission of the SNR models is the result of a series of
  processes.  Gas hit by the blast wave is heated, but at first the
  rise of the temperatures affects only the ions.  Afterwards,
  collisions transfer energy to the electron gas, gradually
  establishing equilibrium between the two fluids.  Since the
  collision frequency is highest in dense regions of the SNR, in
  particular the surfaces of the clouds, the electron temperature
  increases rather inhomogeneously, and, thus, the bremsstrahlung
  emission from the remnant is correlated very clearly to the presence
  and properties of the clouds.  Typically, the early light curves of
  the models exhibit a steep decline in all energy bands as the shock
  wave expands rapidly, leading to cooling of the post-shock matter.
  Once the shock wave starts to interact with the clouds, the decline
  is much slower; some models even enter a plateau phase of
  essentially constant luminosity.  During this phase, which
  corresponds most likely to the current observational appearance of
  \VelaJr, we estimate luminosities in the UV and X-ray bands between
  a few $\zehn{32}$ and a few $\zehn{33} \, \erg / \mathrm{s}$, i.e.,
  of the same order of magnitude as observations suggest.  As noted
  above, the emission reflect the inhomogeneities of the ISM, yielding
  patchy emissivity maps in which the clouds are connected by
  filamentary structures, similar to what can be seen in observations.
  After the passage of the clouds by the shock wave, the effects of
  the interaction tend to become less prominent compared to the
  expansion, and the luminosities decrease again, though slower than
  in the initial stages.  Models with high explosion energies per unit
  mass of the ejecta tend to produce an emission at the upper bound of
  luminosities given above; the same tendency can be observed if we
  increase the ISM density or add a large number of small clouds in
  the vicinity of the explosion.  The light curves show pronounced
  differences across the range of photon energies.  As the SNR expands
  and the mean temperatures drop, the emission gradually shifts to
  lower frequencies.  High-energy bands fade away very quickly, while
  the bands we are most interested in, UV and low-energy X-ray, remain
  strong during the entire evolution.  In most models, the X-ray band
  of 1 - 7 keV is more luminous than the UV band of 80-120 eV by a
  factor of a few.  Exceptions from this behaviour can be found for
  spherical and low-energy explosions, as well as a very dense ISM
  with many clouds.
\item Using (passively advected) tracer particles we follow the
  evolution of three elements.  Based on the results by
  \cite{Maeda_Nomoto__2003__apj__Bipolar_SN:Nucleosynthesis_and_Implications_for_Abundances_in_EMP_Stars},
  we assume that titanium and nickel are present only in the bipolar
  outflow, while oxygen is distributed across all angles of the
  explosion.  During the expansion of the ejecta, all elements are
  redistributed to lower velocities. Their velocity spectra broaden
  considerably when the flow of the post-shock matter past the clouds
  decays by instabilities into small scale structures, and particles
  are trapped inside the SNR at low velocities.  Based on the
  evolution of the velocity spectra, we can distinguish a spherical
  explosion from a bipolar one: the latter exhibits a broader spectrum
  and shows nickel at much higher velocities.
\end{enumerate}

The main caveat for a comparison of our models with observations is
our limited modelling of the radiation processes.  The observations of
the Vela Jr.~SNR show large contributions due to non-thermal processes
and thermal line emission, which we currently do not include.  We
tried to assess the importance of this shortcoming by applying a
criterion for X-ray synchrotron emission
\citep{Vink__2012__aapr__Supernovaremnants:theX-rayperspective}.  We
do indeed expect that our models would be strong sources of
non-thermal synchrotron radiation.  However, in regions where the
shock hits the major clouds, X-ray synchrotron emission seems to be
suppressed due to the low shock velocity unless the explosion was very
energetic.

The main difficulty in identifying a possible progenitor star is that
most of the observational properties of the SNR are affected by the
characteristics of the ISM, in particular by the distribution of
clumps and clouds, and by foreground absorption.  The initial
conditions of the explosion are washed out to a high degree during the
interaction of the shock wave with the ISM.

Based on our results, we conclude that the current appearance of the
Vela Jr.\ SNR is compatible with most of our simulations.  Though the
differences between our models do not allow us to determine the
progenitor of the SN with certainty, we can at least point towards the
most likely combinations of explosion properties and ages of the SNR.
The assumption of a hypernova around the $12^{\mathrm{th}}$ century
(with a considerable uncertainty that may be up to a few hundred years
) and a less energetic explosion a few thousand years ago agree well
with the observations.  Hence, observations constraining either the
explosion energy or the age of the remnant would constrain the other
of the two parameters.  Confronting results of our simulations with
additional observations of the Vela Jr. SNR and of its environment in
radio, optical, UV, and X- and gamma-rays can provide valuable
information on the SNR progenitor type.  Given the current set of
models, observations identifying the thermal contribution of the
emission would be most helpful, while the inclusion of non-thermal
processes and line emission is the most important requirement for
future modelling.  It would also help us to better constrain the
expansion rate and the age of the SNR, and its distance.

\section{Acknowledgements}
\label{Sek:Ackno}

AFI and GFS were partially supported through the Grant of RF
``11.G34.31.0076''.  AFI acknowledges discussions and joint work with
N.~Chugai and Yu.~V.~Pakhomov related to the optical properties of
Vela Jr., and thanks V.~Burwitz, K.~Dennerl and F.~Haberl for their
contribution to the X-ray imaging of \VelaJr, and especially Bernd
Aschenbach for his unwaivering support of the whole effort to prove
the young age of the remnant.  MO acknowledges discussions with
M.A.~Aloy, D.~Patnaude, R.~Fesen, and D.~Milisavlejic as well as
support from the European Research Council (grant CAMAP-259276), and
from the Spanish Ministerio de Ciencia e Innovaci{\'o}n (grant
AYA2010-21097-C03-01 \emph{Astrof{\'i}sica Relativista
  Computacional}).  We are grateful for technical support by the
system administrators of the Universitat de Val{\`e}ncia, in
particular C.~Aloy.  We thank the anonymous referee for his/her
valuable comments.

% \bibliographystyle{mn2e}
% \bibliography{./biblio}

\end{document}